\documentclass[a4paper,10pt]{scrartcl}
\usepackage[utf8x]{inputenc}
\usepackage{color}
\usepackage{psfrag}

\usepackage{amssymb,amstext,amsmath,amsthm,amsxtra}

\newcommand{\bea}{\begin{eqnarray}}
\newcommand{\eea}{\end{eqnarray}}
\newcommand{\nn}{\nonumber}
\newcommand{\be}{\begin{equation}}
\newcommand{\ee}{\end{equation}}

\theoremstyle{definition}

\theoremstyle{plain}
\newcommand{\HH}{\mathcal{H}}

\newcommand{\CC}{\mathcal{C}}

\newcommand{\id}{\mathbf{1}}
\DeclareMathOperator{\tr}{tr}

\newcommand{\lalg}[1]{\mathfrak{#1}}  % Lie algebra
\newcommand{\SU}{\mathrm{SU}}

\newcommand{\U}{\mathrm{U}}
\newcommand{\Spin}{\mathrm{Spin}}

\newcommand{\su}{\lalg{su}}

\newcommand{\dd}{\mathrm{d}}

\usepackage{graphicx}

\newcommand{\q}{\quad}

\newcommand{\SixJ}[6]{\left\{\begin{array}{ccc}#1\,&#2\,&#3\,\\
#4\,&#5\,&#6\,\end{array}\right\}}

\title{Holonomy Spin Foam Models: Definition and Coarse Graining}
\author{Benjamin Bahr$^1$, Bianca Dittrich$^{2,3}$,\\ Frank Hellmann$^3$, Wojciech Kaminski$^3$\\
\small $^1$ DAMTP, University of Cambridge,
\small  Wilberforce Road, Cambridge CB3 0WA, UK\\
\small $^2$ Perimeter Institute, %f. Theoretical Physics,
\small  31 Caroline St. N, Waterloo, ON N2L 2Y5, Canada\\
\small   $^3$ MPI f. Gravitational Physics, %Albert Einstein Institute,
 \small Am M\"uhlenberg 1, D-14476 Potsdam, Germany
 }

\date{}

\begin{document}

\maketitle

\begin{abstract}
We propose a new holonomy formulation for spin foams, which naturally extends the theory space of lattice gauge theories. This allows current spin foam models to be defined on arbitrary two--complexes as well as to generalize current spin foam models to arbitrary, in particular finite groups.  The similarity with standard lattice gauge theories allows to apply standard coarse graining methods , which for finite groups can now be easily considered numerically. We will summarize other holonomy and spin network formulations of spin foams and group field theories and explain how the different representations arise through variable transformations in the partition function. A companion paper will provide a description of boundary Hilbert spaces as well as a canonical dynamic encoded in transfer operators.

 \end{abstract}

\section{Introduction}

Spin foam models are non-perturbative candidate theories for quantum gravity \cite{bookcarlo, alexreview}, which succeed in combining discreteness of spectra of geometric operators \cite{smolin,ash} with the correct semi-classical behaviour of its elementary building blocks \cite{Barrett:1998gs, Barrett:2010ex,Conrady:2008mk}. One of the key challenges is to obtain a better grasp on the continuum limit of spin foams, obtained by considering the limit of (infinitely) many building blocks. This will not only answer questions on the large scale behaviour (which here we will understand as the one involving many building blocks) but also address the status of diffeomorphism symmetry in these models \cite{Dittrich:2008pw,Bahr:2009ku,Bahr:2009qc,Rovelli:2011fk,DittrichZak}.

To investigate this kind of limit involving many building blocks renormalization and coarse graining methods seem to be most appropriate \cite{fotini,oeckl,Dittrich:2011zh,Dittrich:2011av}. To this end an understanding of the space of models, on which the renormalization flow takes place, is crucial. This allows also also to distinguish  between  relevant and irrelevant choices in the construction of spin foam models and helps to reveal the basic dynamical mechanisms of these models.

In this work we therefore propose a unified description of spin foam models based on group or holonomy variables. In this form spin foam models appear to be similar to lattice gauge theories, have however a richer structure of dynamical ingredients. In the description with the fewest parameters our models are defined by an edge function $E$ and a face weight $\omega$, whereas lattice gauge theories just feature the face weights $\omega$. All the current spin foam models, namely Barrett--Crane, EPRL and FK models \cite{Barrett:1997gw,Engle:2007wy,Freidel:2007py}, can be described by this choice.

For this description to be useful for coarse graining one needs to show that these structures are stable under coarse graining (within some approximation scheme). Indeed we will investigate different coarse graining schemes for two and three dimensional models analytically and numerically and show that this space of models presented here supports a rich structure for the renormalization flow. To be able to apply numerical techniques we consider spin foam models with finite groups, a concept introduced in \cite{Bahr:2011yc,Dittrich:2011zh}. A further advantage of the holonomy description here is that models, analogous to the Barrett Crane or EPRL model, can be easily introduced also for finite groups. This will ease the investigation of (for instance symmetry) properties of these models as well as the numerical investigation of the large scale limit.

There are different holonomy representations proposed already in the literature \cite{PFEIF,val,Magliaro:2010ih,benloops}, the relation to these will be explained in section \ref{rela}. The advantage of the formulation presented in this work is that it features -- compared to the other formulations -- a minimal set of model parameters and makes maximal use of the common properties of the spin foam models proposed so far in the literature. Most prominently that the imposition of the simplicity constraints factorizes in a specific sense, see also \cite{alexreview}.

We will also explain the relation to the so--called operator spin foam models \cite{Bahr:2010bs}, which share with the models proposed here, that they are defined on arbitrary two--complexes. Indeed the structure of the holonomy representation worked out here makes the definition on a two--complex completely natural. %Furthermore this description allows easily a generalization of a given model to finite groups \cite{Bahr:2011yc}. %Obtaining in this way a family of simplified models might facilitate the (numerical) investigations of the large scale limit, see \cite{Dittrich:2011zh,Dittrich:2011av}.

%This holonomy formulation also allows a natural definition of boundary Hilbert spaces that will be described in section . This gives a unified description of a universal boundary space for all current spin foam models. Also we will see how different choices of choosing the boundary correspond to different Hilbert space description, including the Hilbert space of projected spin networks \cite{Alexandrov:2002xc,Livine:2002ak}. Using the latter we can easily re--obtain spin foams in the usual form, i.e. with vertex weights evaluated in the spin network basis.

Other advantages of the holonomy representations introduced here are the natural definition of boundary Hilbert spaces and a straightforward relation to projected spin networks \cite{Alexandrov:2002xc,Livine:2002ak} as well as allowing the investigation of the semi--classical limit for arbitrary complexes by exploring the distributional character of the integration kernels of the partition functions. These features will be explained in companion papers \cite{HS, SL}.

~\\

In section \ref{sec-HolForm} we give the first main result of this paper, a definition of spin foam models as generalized lattice gauge theories written entirely in terms of holonomies. Section \ref{rela} discusses the relation to operator spin foams and other holonomy formulations for spin foams, including group field theories. The next section gives a description of the Barrett--Crane and EPRL  spin foam models in the holonomy formulation. These insights are used to generalize the Barrett Crane and EPRL models to finite and other Lie groups.

In section \ref{coarse} we discuss coarse graining of these models, in particular in two and three dimensions. We close with an outlook and summary in sections \ref{out} and \ref{sum}.

\section{A short motivation of the models}\label{motivation}

Here we shortly motivate some ideas behind the holonomy representation for spin foams, which will be explained in more detail later on.

Let us start with reviewing standard lattice gauge theory. Such a theory requires the choice of a gauge group $G$ and face weights $\omega$, which are class functions on the group. These face weights encode all the dynamics of the given model, as the partition function is given by
\bea\label{eins}
Z=\int \left(\prod_e \dd g_e\right)  \left( \prod_f \omega(g_e g_{e'} \cdots)\right) \quad .
\eea
On the other hand the face weights for i.e. lattice Yang Mills theory are defined using the background metric of the lattice. This should however be avoided if one wants to quantize gravity, where the metric itself is a variable. Indeed, the starting point for spin foam model construction is BF theory where the face weights are given by $\omega=\delta_G$ the delta function on the group. This makes the partition function in general divergent (even for compact groups but not for finite ones). This divergence indicates another gauge invariance, known as translation symmetry. BF theory is a topological field theory, which coincides with the zero coupling or zero temperature limit of lattice Yang Mills theory.

To construct a non--topological theory and nevertheless keep the trivial face weights, we have to introduce more structure. First let us just double the variables from one group element $g_{e}$ per edge to one group element $g_{ve}$ per edge--vertex pair (or half edge). We define $g_{ev}=(g_{ve})^{-1}$.  Then the partition function (\ref{eins}) reads
\bea\label{zwei}
Z=\int \left(\prod_{(ve)} \dd g_{ve}\right)  \left( \prod_f \omega(g_{ve} g_{ev'} g_{v'e'} \cdots)\right) \quad
\eea
and we can regain the form (\ref{eins}) by  variable redefinitions $g_e=g_{ve}g_{ev'}$  and performing $|E|$ trivial integrations. Here and in the following we assume a normalized and left and right invariant group integration measure.

To actually change the model we introduce another set of variables, namely a group element $g_{ef}$ per face--edge pair.  These are inserted into the face holonomies so that $\omega_f$ is now evaluated on $g_f=g_{ve}g_{ef} g_{ev'} g_{v'e'} \cdots$.  In this way the alternative face holonomy  $\tilde G_f=g_{ve}g_{ev'}g_{v'e'}\cdots$ can nevertheless take non--trivial values even if we choose $\omega_f=\delta_G$. Allowing all possible $g_{ef}$ would however again trivialize the model. (This time it  corresponds to the high temperature fixed point of lattice Yang Mills, where $\omega_f \equiv 1$ is the constant function.)  A non-trivial model can be obtained by weighing the inserted group elements $g_{ef}$ via a function $E$. The final model will be given by
\be\label{drei}
Z'=\int \left(\prod_{(ve)} \dd g_{ve}\right)     \left(\prod_{(fe)} \dd g_{fe}\right)   \left(\prod_{(fe)} E(g_{ef})\right)    \left( \prod_f \omega_f(g_{ve}g_{ef} g_{ev'} g_{v'e'} g_{e'f}\cdots)\right) .
\ee

As will be explained in the following the partition functions of all current spin foam models can be expressed in this way. The choice of the function $E$ determines the model, the face weights are usually fixed to be $\delta_G$.

The models can be rewritten in several ways, we will present different formulations in section \ref{rela}. Here we will just note that one could integrate out the $g_{ef}$ variables in the formulation (\ref{drei}) to obtain effective face weights
\bea\label{vier}
&&\omega'_f(g_{ev}g_{ve'},g_{e'v'}g_{v'e''},\cdots)= \nn\\
&&\q\q\q\int    \left(\prod_{(fe)} \dd g_{fe}\right)   \left(\prod_{(fe)} E(g_{ef})\right)    \left( \prod_f \omega_f( g_{ev} g_{ve'} g_{e'f}g_{e'v'}g_{v'e''}g_{e''f}\cdots)\right)\q.\q\q
\eea
Such a formulation (for the Barrett Crane model) appeared for instance in \cite{PFEIF,val} and is nearest to standard lattice gauge theory. See also \cite{PFEIF,val2,aa} for a geometric interpretation of the holonomy variables and simplicity constraints involved there. 

Note that  the face weights are not just class functions of a face holonomy, as is the case in standard lattice gauge theory, but will be functions of a certain number of group elements.

\section{Holonomy Formulation of Spin Foam Models}\label{sec-HolForm}

Let us describe the models in more detail. These will be defined on an oriented $2$-complex $\mathcal{C}$ with vertices $v$, edges $e$ and faces $f$. The gauge group will be denoted $G$, which we will assume to be finite or compact  and equipped with a normalized left and right invariant group (Haar) integration measure.

\begin{figure}[hbt]
\begin{center}
	\psfrag{v}{$v$}
	\psfrag{vd}{$v'$}
	\psfrag{e}{$e$}
	\psfrag{ed}{}
	\psfrag{f}{$f$}
	\psfrag{gef}{$g_{ef}$}
    \psfrag{gev}{$g_{ve}$}
\includegraphics[scale=0.6]{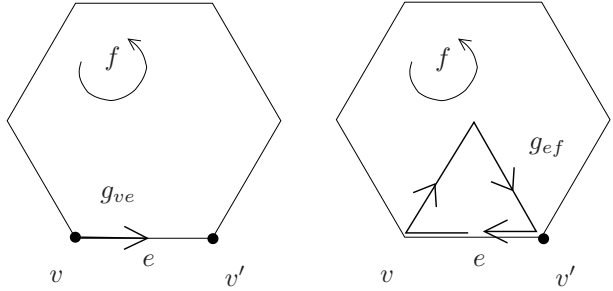}
\caption{\label{Fig:VariableGevGef}Holonomy variables associated to half edges ($g_{ve}$) and  wedges ($g_{ef}$) of a face.}
\end{center}
\end{figure}

The variables in which the partition function will be expressed are group elements $g_{ef}$ assigned to pairs $e\subset f$ and $g_{ve}$ for $v\subset e$ with the understanding that $g_{ev}=g_{ve}^{-1}$. One can identify them as holonomies of a $G$-connection along certain paths within faces $f$ (see Figure \ref{Fig:VariableGevGef})

From these data we can define the holonomy $G_f$
\begin{eqnarray}\label{Eq:TrivialPath}
G_f\;=\;g_{ve}g_{ef}g_{ev'}g_{v'e'}\cdots g_{e^{(n)}v}.
\end{eqnarray}
\noindent With the help of Figure \ref{Fig:VariableGevGef}, one can readily see that $G_f$ describes the holonomy associated to a \emph{geometrically} trivial path, i.e.~the choice $\omega=\delta_G$ as a weight to this holonomy appears most natural and is also used in the EPRL-FK or BC model. For regularization purposes, this $\delta_G$ is often replaced by a heat kernel, see e.g.~\cite{Bonzom:2010ar}. We will therefore leave the face weight  $\omega$ as a parameter of the theory. As in standard lattice gauge theory $\omega$ is required to be a class function of the group, furthermore we ask that $\omega(g)=\omega(g^{-1})$ is satisfied for all $g\in G$.

The class of models we are considering are specified further by a subgroup $G' \subset G$. We will choose the edge function $E$ to be invariant under conjugation with elements in  $G'$. From (\ref{vier}) one can see that $E$ functions which are invariant under conjugation of the full group, i.e. class functions, will lead to effective face weights which can be considered to just depend on the (reduced) face holonomy $g_{ev}g_{ve'}g_{e'v'}\cdots$. This case will therefore lead to a model equivalent to a lattice gauge theory (with face weights $\omega_f$ that might explicitly depend on the face, more precisely on the number of edges in the face).  In this sense the choice of $G'$ defines an invariance property of the effective face weights (\ref{vier}).
As for the face weights  we will  furthermore require  from $E$ the property $E(g)=E(g^{-1})$ for $g\in G$.

%{\red might include reality of E and omega function}

Finally the state sum is defined as
\begin{eqnarray}\label{Eq:HolonomyStateSum}
Z\;=\;\int dg_{ev} dg_{ef}\;\prod_f\omega(G_f)\,\prod_{ef} E(g_{ef}) \q .
\end{eqnarray}

For the choice $\omega=\delta_G$ the partition function may be divergent, even for compact groups. The definition is extendable to non--compact groups, but in this case there is an ambiguity in the normalization of the Haar measure and additional divergences due to gauge orbits.
However, for finite groups the delta distribution is just a regular functions, thus the above is well defined.

\subsection{Parametrization of models}\label{par}

The space of such holonomy spin foam models is described by a choice of the edge functions  $E$ and the face weights $\omega_f$. In the following we will give the spin representation of these entities as such a representation also offers a convenient parameterization of the space of models.

The face weight $\omega_f$ is a class function and as usual can be expanded as

\be\label{eq-faceDistrib}
\omega_f(g) = \sum_{\rho} \dim(\rho)\, \omega^\rho \chi_\rho(g),
\ee
where $\chi_\rho=  \tr_{\rho}\left(D_\rho(g)\right)$ is the character of the unitary irreducible representation (irrep) $\rho$ with $D_\rho(g)$ the representation matrix of $g$.

%Further note that as the conditions on $E$ and $\omega_f$ are linear, the space of theories given two groups $G' \subset G$ carries a linear structure. We will often set $\omega_f = \delta_G$, or $\omega^\rho = 1$ in which case this space is simply parametrized by the functions $E$ satisfying the conditions in section  \ref{sec-HolForm}. We call this space $\EE(G,G')$.

The edge functions  can be expanded in terms of the irreducible unitary representations of $G$ and $G'$, which we denote $\rho$ and $k$ respectively. To this end  we utilize  injection maps $I(\rho, k)_d$:
\be \label{eq-GeneralInjection}
I(\rho, k)_d \in Hom_{G'}(\bar\rho, k),
\ee
where the $d$ index labels a basis of $Hom_{G'}(\bar\rho, k)$ the space of $G'$ covariant function  %$k$ and $\rho^*$, the dual of $\rho$.
from the representation space of $\bar\rho$, the dual of $\rho$, to the $G'$ irrep $k$. 

The map $I(\rho, k)_d$  identifies the  representation $k$, which in case there are more than one representation is furthermore labelled by $d$,  in the reduction of the irrep $\rho$ over the group $G'$. The index $d$ is hence only non--trivial in cases where there are more than one $k$ representations included in $\rho$.

We assume that the basis is normalized (up to a phase, not relevant in the following)
\be \label{eq-GenInNormalisation}
I(\rho, k)_d^\dagger I(\rho, k)_{d'} = \delta_{dd'} \id_k.
\ee

This allows to  expand $E$ as
\be\label{Eq:DecompositionOfEFunctions}
E(h) = \sum_{\rho, k} \dim(\rho) e^{\rho}_{k,dd'} \tr_\rho\left(D_\rho(h)\, I(\rho, k)_d {I}^\dagger(\rho, k)_{d'}\right)\q .
\ee
Hence for given groups $G,G'$  the choice of $E$ function can be encoded into the parameters $e^\rho_{k,dd'}$. The index $d$ is a degeneracy index for the case where the $G'$ reducible representation $\rho$ contains more than one copy of the irrep $k$. For $G=SU(2) \times SU(2)$ and $G'=SU(2)_{diag}$ this case does not occur, and we have only to deal with the set $e^\rho_k$.

$E$ encodes the analogue of the simplicity constraints, arising in the Plebanski formulation of gravity, for the spin foam model at hand, see for instance \cite{alexreview}. As in the classical Plebanski formulation, where the simplicity constraints are enforced with Lagrange multipliers, we had to introduce here additional variables, the wedge holonomies $g_{ef}$.

If the face weight is given by the delta function, the product of group elements around the face, $G_f$, is flat. This is however not the usual holonomy around the face $\tilde{G}_f=g_{ve}g_{ev'}g_{v'e'}\cdots$, but $\tilde{G}_f$ interwoven with the $g_{ef}$. If we force $g_{ef} = 1$ by choosing $E=\delta_G$ we have $G_f = \tilde{G}_f$, and obtain a theory of flat connections, that is $BF$ theory. Choosing a different $E$ function relaxes the constraints on flatness, just as we expect from the simplicity constraints.

%We will see the precise relationship between the simplicity operators in the usual operator spin foam models and the functions $E$ in the next section.
%

\subsection{Boundaries and Hilbert spaces}\label{hilb}

We can also include boundaries into our models, which are represented as graphs $\Gamma$ in the two--complex $\CC$.
The set of edges  of $\Gamma$ will be denoted by $\Gamma_e \subset \CC_e$ and the set of vertices $\Gamma_v \subset \CC_v$. We  write $\Gamma_{ev}$ for the set of pairs $v \in e$ in $\Gamma_v \times \Gamma_e$.

By dropping integration over the group elements associated to the pairs of $\Gamma_{ev}$ we arrive at the partition function 
\be\label{eq-PartFunc}
Z^\Gamma (\CC)[g_{ev}] = \int \left(\prod_{e \subset f}  \dd g_{ef}\right) \left( \prod_{\substack{v \subset e \\ef \notin \Gamma_{ev}}} \dd g_{ev} \right) \left(\prod_{e \subset f} E(g_{ef})\right) \left(\prod_f \omega_f(g_f)\right).
\ee
which can now be interpreted as an element of the Hilbert space $L^2(G^{|\Gamma_{ev}|})$. (Integrability in this general form is only guaranteed for finite groups).

As the integrand of the partition function is invariant under
$
g_{ev} \rightarrow {h'_e}^{-1}g_{ev}g_v,\,\,
h_{ef} \rightarrow {h'_e}^{-1}h_{ef}h'_e
$
for $h'_e \in G' \subset G$ and $g_v \in G$,
 the partition function can  be considered to live in a smaller subspace of $L^2(G^{|\Gamma_{ev}|})$, that is, \be Z \in \HH^\Gamma = L^2\left(G^{|\Gamma_{ev}|}{\Bigl /}{\left(G^{|\Gamma_v|} \times {G'}^{|\Gamma_e|}\right)}\right).\ee
This Hilbert space also describes the effective face weights $\omega'$ defined in (\ref{vier}).

We call this space the universal boundary space for the holonomy spin foams, which is fixed by a choice of graph, and the two groups $G,G'$. This Hilbert space and its relation to projected spin networks \cite{Alexandrov:2002xc,Livine:2002ak} will be further explored in the companion paper \cite{HS}.   %and write $\HH^\Gamma_{UBS}$.

\section{Relation to other formalisms}\label{rela}

The form of the state sum (\ref{Eq:HolonomyStateSum}) was chosen such that it mimics the main features of the EPRL (and other known) spin foam models, but has a minimal set of free parameters. This appears to be a very convenient choice, but it is by far not the only one. There have been other forms used in the literature, see for instance \cite{PFEIF,val,Magliaro:2010ih,benloops}, which include weights for different geometric objects, which might appear under coarse graining. The aim of this section is to give a precise relation between the $E$-function formalism used in this paper, and the other formulations.

\subsection{Operator Spin Foam models}\label{osfm}

The operator spin foam models (OSFM), introduced in \cite{Bahr:2010bs}, were originally formulated with the aim of allowing for arbitrary operators $P_e$ associated to edges $e$ of $\mathcal{C}$. The state sum $Z$ is evaluated by contracting the $P_e$ at the vertices $v$ of $\mathcal{C}$ according to the combinatorics of the $2$-complex.

Starting with a $2$-complex with irreps $\rho_f$ of $G$ associated to faces $f$, an OSFM is specified by operators mapping the intertwiner space onto itself
\begin{eqnarray}\label{Eq:OSFMOperator}
P_e\;:\;\text{Inv}_G(\rho_{f_1}\otimes\ldots\rho_{f_n})\;\longrightarrow\;\text{Inv}_G(\rho_{f_1}\otimes\ldots\rho_{f_n}),
\end{eqnarray}
\noindent as well as face weights $\omega^{\rho_f}$. There is a dual formulation of this model \cite{benloops}, using holonomies $h_{ef}\in G$ for all pairs $(e\subset f)$. This holonomy can be thought of as starting at one vertex $v$, going to the middle of the  face $f$, and going to another vertex $v'$, so that $e$ goes from $v$ to $v'$ in the orientation induced by $f$.
\begin{figure}[hbt]
\begin{center}
	\psfrag{v}{$v$}
	\psfrag{vd}{$v'$}
	\psfrag{e}{$e$}
	\psfrag{ed}{}
	\psfrag{f}{$f$}
	\psfrag{hef}{$h_{ef}$}
\includegraphics[scale=0.4]{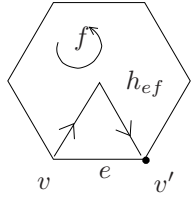}
\caption{\label{Fig:VariableHef} Holonomy variables in the dual description of operator spin foam models.}
\end{center}
\end{figure}
\noindent The Fourier transform of the $P_e$ operators is given by functions $C_e$
\begin{eqnarray}\label{Eq:OSFMCFunctions}
C_e(h_1, \ldots, h_n)\;=\;\sum_{\rho_k}\;\left(\prod_k\,\dim\rho_k\,   D_{\bar\rho_k}(h_k)_{m_k}{}^{n_k}\right)\;(P_e)^{m_1\ldots m_n}{}_{n_1\ldots n_n}.
\end{eqnarray}

\noindent where we assume for simplicity that the edge orientation agrees with all orientations from all faces meeting at it. %The $\tilde\omega_{\rho_f}$  are the Fourier coefficients of a class function $\omega$ associated to the faces. 
Then, the state sum in terms of holonomies is given by
\begin{eqnarray}\label{Eq:OSFMHolonomyStateSum}
Z_{\rm OSFM}\;=\;\int dh_{ef}\;\left(\prod_f \omega(G_f)\right)\prod_e C_e(h_{ef_1}, \ldots, h_{ef_n})
\end{eqnarray}

\noindent The $G_f$ is the ordered product of holonomies $h_{ef}$ around a face $f$. This forms the same path as (\ref{Eq:TrivialPath}).

The $E$-function formalism is a special case of the OSFM, in which the edge operator $P_e$ factorizes over the edges, and the invariance property $C_e(h_1,\ldots, h_n)=C_e(g_Lh_1g_R,\ldots, g_Lh_ng_R)$, which follows from (\ref{Eq:OSFMOperator}) and (\ref{Eq:OSFMCFunctions}), is explicitly enforced by projectors onto the intertwiner space.
One can write the $E$-function state sum (\ref{Eq:HolonomyStateSum}) in the form (\ref{Eq:OSFMHolonomyStateSum}) by performing the variable transformation
\begin{eqnarray}
h_{ef}\;=\;g_{ve}\,g_{ef}\,g_{ev'}.
\end{eqnarray}

\noindent This removes the $g_{ev}$ - dependence of the face functions $\omega$, and one can carry out the integration over the $g_{ev}$, defining
\begin{eqnarray}\label{GA}
C_e(h_{ef_1},\ldots,h_{ef_n})\;:=\;\int_{G^2} dg_{ev}dg_{ev'}\prod_{f_k\supset e}\,E(g_{ev}h_{ef_k}g_{v'e})
\end{eqnarray}

\noindent for each edge. This results in (\ref{Eq:OSFMHolonomyStateSum}), as one can readily check.

\begin{figure}[hbt]
\begin{center}
	\psfrag{P}{$P_e$}
	\psfrag{E}{$\tilde E$}
	\psfrag{=}{$=$}
	\psfrag{d}{$\vdots$}
	\psfrag{f}{$f$}
	\psfrag{hef}{$h_{ef}$}
\includegraphics[scale=0.6]{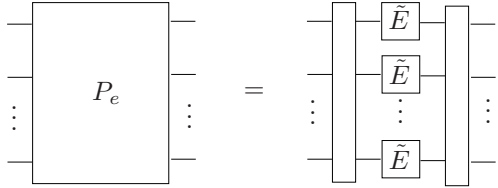}
\caption{Graphical notation of the relation between the OSFM version of the edge propagator $P_e$ and the one used in (\ref{Eq:HolonomyStateSum}), which is also used in the  BC and EPRL models. }\label{Fig:RelationOSFM-EFunctions}
\end{center}
\end{figure}

Hence one can see that the $E$-function state-sum (\ref{Eq:HolonomyStateSum}) is a special case of the OSFM state-sum (\ref{Eq:OSFMHolonomyStateSum}), in which the operator $P_e$, which is a map from the invariant subspace of $\rho_{f_1}\otimes \cdots \rho_{f_n}$ to itself, can be written as a part which factorizes over every face, and where the right- and left-invariance are enforced by sandwiching the product between two Haar intertwiners. This Haar projector is just the group averaging in (\ref{GA}) translated into the spin representation. The factorizing property means that $P_e$ can be written in the following way
\bea
P_e \,=\, P_{Haar} \cdot P_e' \cdot P_{Haar} \,=\, P_{Haar}\cdot \left( \tilde E_{\rho_1} \otimes \tilde E_{\rho_2} \otimes \cdots \tilde E_{\rho_n}\right)\cdot P_{Haar} \q ,
\eea
which again is (\ref{GA})  group--Fourier transformed (see figure \ref{Fig:RelationOSFM-EFunctions}). Here $\tilde E_{\rho_i}$ are (usually not covariant under the group action) maps acting on $\rho_i$ and with matrix components given by the group Fourier transform of the functions $E$, i.e.
\begin{eqnarray}
(\tilde{E}_\rho )^{m}{}_{n}\;=\;\sum_{d,d',i}e^{\rho}_{k\,dd'}\;I(\rho,k)^m{}_i(I^\dagger(\rho,k))^i{}_n
\end{eqnarray}
\noindent using the notation from (\ref{Eq:DecompositionOfEFunctions}).

\subsection{Group Field Theory formulation}

\begin{figure}[hbt]
\begin{center}
	\psfrag{v}{$v$}
	\psfrag{vd}{}
	\psfrag{e}{$e$}
	\psfrag{ed}{$e'$}
	\psfrag{f}{$f$}
	\psfrag{gvfe}{$g_{vf}^{e'}$}
	\psfrag{gvfed}{$g_{vf}^{e}$}
\includegraphics[scale=0.4]{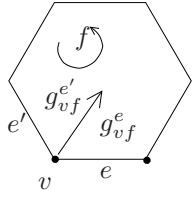}
\caption{ \label{Fig:VariableGvfe}  Variables in the group field theory formulation.}
\end{center}
\end{figure}

The Group Field Theory (GFT) formulation \cite{DePietri:1999bx,Reisenberger:2000zc,Geloun:2010vj,oritigft} of the spin foam state sum is based on a quantum theory of fields defined on the group $G$, such that the Feynman graphs are Spin Foam amplitudes associated to $2$-complexes\footnote{Where the specific form of the action is designed to control the class of appearing $2$-complexes, see e.g.~\cite{Gurau:2009tw}.}. For one of these $2$-complexes the group holonomies are pairs of holonomies $g_{vf}^e, g_{vf}^{e'}$, where $v\subset e, e'\subset f$. Both are interpreted as going from the vertex $v$ to the center of the face $f$, and each is associated with one of the two edges $e$, $e'$ that border $f$ and meet at $v$ (see figure \ref{Fig:VariableGvfe}).

\noindent As usual, we denote the inverse by $(g_{vf}^e)^{-1}=:g_{fv}^e$. A natural class of models arises if one indeed imposes $g_{vf}^e\stackrel{!}{=}g_{vf}^{e'}$, as Figure \ref{Fig:VariableGvfe} suggests, but the GFT models allow for relaxation of this condition.

The ingredients for the GFT amplitude associated to $\mathcal{C}$ are edge propagator functions $\Pi_e$, each depending on all $g_{fv}^e$ for a given $e$, and vertex functions $\mathcal{V}_v$, each depending on all $g_{vf}^e$ for a given $v$. The functions have to satisfy the following symmetries:
\begin{eqnarray}\label{Eq:GFTSymmetries}
\mathcal{V}_v(\ldots,g_{vf}^e, g_{vf}^{e'},\ldots)\;&=&\;\mathcal{V}_v(\ldots,g g_{vf}^e, g g_{vf}^{e'},\ldots)\\[5pt]\nonumber
\Pi_e(\ldots,g_{vf}^e, g_{v'f}^{e},\ldots)\;&=&\;\Pi_e(\ldots,g_{vf}^e g, g_{v'f}^{e} g,\ldots)
\end{eqnarray}
\noindent The symmetries (\ref{Eq:GFTSymmetries}) make sure that $\mathcal{V}$ effectively only depends on the combinations $g_{fv}^eg_{vf}^{e'}$, and the propagator functions $\Pi$ depend only on $g_{vf}^eg_{fv'}^{e}$. Also, both functions $\mathcal{V}_v$ and $\Pi_e$ have to be gauge-invariant with respect to gauging at an edge, i.e.~$g_{vf}^e\to g\,g_{vf}^e$ for fixed $v, e$, and all $f$ which appear as variables in the respective function. As a result, the $\Pi_e$ will be invariant under left or right shift of its arguments. The $\mathcal{V}_v$ can be decomposed into spin network functions of the boundary graph of the vertex $v$, which gives the usual spin representation of the vertex amplitudes.  %see e.g.~the construction in (...).

The state sum for the GFT models is given by
\begin{eqnarray}\label{Eq:GFTHolonomyStateSum}
Z_{\rm GFT}\;=\;\int dg_{vf}^e\;\prod_e \Pi_e\big(g_{vf_1}^e,\ldots,g_{v'f_n}^e\big)\;\prod_v \mathcal{V}_v\big(g_{vf_1}^{e_1},\ldots, g_{vf_n}^{e_m}\big)
\end{eqnarray}

\noindent The $E$-function form (\ref{Eq:HolonomyStateSum}) for the state sum $Z$ can be written as a GFT state sum (\ref{Eq:GFTHolonomyStateSum}) for $\omega =\delta$. If the product of the $h_{ef}$ around the faces are constrained to be the unit element, one can write (uniquely up to global right shift of all $g_{vf}^e$ for some fixed $f$ by the same $g\in G$)
\begin{eqnarray}
h_{ef}\;=\;g_{vf}^eg_{fv'}^{e},
\end{eqnarray}
\noindent imposing $g_{vf}^e\stackrel{!}{=}g_{vf}^{e'}$ in the state sum. This leads to
\begin{eqnarray}
Z\;&=&\;\int dg_{vf}^e\;\prod_e C\big(g_{vf}^eg_{fv'}^{e},\ldots \big)\;\prod_{vf}\delta\big(g_{vf}^eg_{fv}^{e'}\big)\\[5pt]\nonumber
&=&\;\int dg_{vf}^e\;\prod_e \;\Pi_e\big(g_{vf_1}^e,\ldots,g_{v'f_n}^e\big)\prod_v \mathcal{V}_v\big(g_{vf_1}^{e_1},\ldots, g_{vf_n}^{e_m}\big)
\end{eqnarray}
\noindent with vertex function
\begin{eqnarray}
\mathcal{V}_v\big(g_{vf_1}^{e_1},\ldots, g_{vf_n}^{e_m}\big)\;:=\;\prod_{f\supset v}\delta\big(g_{vf}^eg_{fv}^{e'}\big)
\end{eqnarray}
\noindent and propagator
\begin{eqnarray}
\Pi_e\big(g_{vf_1}^e,\ldots,g_{v'f_n}^e\big)\;:=\;C\big(g_{vf}^eg_{fv'}^{e},\ldots\big) \q .
\end{eqnarray}

\subsection{Vertex amplitude formulation}

\begin{figure}[hbt]
\begin{center}
	\psfrag{v}{$v$}
	\psfrag{vd}{}
	\psfrag{e}{$e$}
	\psfrag{ed}{$e'$}
	\psfrag{f}{$f$}
	\psfrag{geef}{$g_{v}^{f}$}
\includegraphics[scale=0.4]{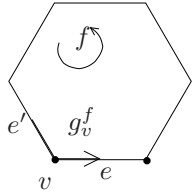}
\caption{\label{Fig:VariableGeef}  Holonomy variables in the vertex amplitude formulation.}
\end{center}
\end{figure}

A formulation closely related to the GFT formalism can be given in terms of holonomies labelled by wedges, i.e.~by a face $f$, and pairs $e,e'$ that meet at a common vertex (see Figure \ref{Fig:VariableGeef}). The holonomy can be thought of as going from the center of $e'$ to the vertex $v$, and from there to the center of $e$.

\noindent One can arrive at a formalism using the $g_{v}^f$ by imposing $g_{e'v}g_{ve}\stackrel{!}{=}g_{v}^f$ (as suggested by figures \ref{Fig:VariableGevGef} and \ref{Fig:VariableGeef}) in the state sum (\ref{Eq:HolonomyStateSum}) and integrating out the gauge holonomies $g_{ev}$, arriving at
\begin{eqnarray}\label{Eq:VertexFunctionHolonomyStateSum}
Z_\text{Vertex}\;=\;\int dg_{ef}dg_v^f \prod_f\omega(G_f)\prod_{ef} E(g_{ef})\prod_v W_v(g_{v}^{f_1},\ldots, g_{v}^{f_n})
\end{eqnarray}

\noindent where $G_f=g_v^fg_{ef}g_{v'}^f\cdots g_{e'f}$ is the ordered product of group elements around a face $f$, according to its orientation. The vertex function $W_v$ for a vertex $v$ is given by
\begin{eqnarray}\label{Eq:VertexFunction}
W_v(g_{v}^{f_1},\ldots, g_{v}^{f_n})\;=\;\int dg_{ev}\prod_{f\supset v}\delta\big(g_{e'v}g_{ve}(g_v^f)^{-1}\big)   \q .
\end{eqnarray}

\noindent The vertex functions $W_v$ have a symmetry that corresponds to gauge transformation associated to edges, i.e.~for an edge $e$ beginning at vertex $v$, $V_v$ is invariant under the transformation of its arguments
\begin{eqnarray}
g_v^f\;\rightarrow\;\left\{\begin{array}{ll}g_v^fh_e &\text{if }f\supset e\\[5pt]g_v^f&\text{if }f\not\supset e\end{array}\right.
\end{eqnarray}
\noindent This invariance allows to interpret $W_v$ as a $G$-spin network function for the boundary graph of the neighbouring $2$-complex of the vertex $v$.
A very similar formulation that uses $G'$-spin networks instead was given in \cite{Magliaro:2010ih}.
\\[5pt]

Writing the state sum model in the form (\ref{Eq:VertexFunctionHolonomyStateSum}) allows for generalizations of the model, by relaxing the condition (\ref{Eq:VertexFunction}), allowing for more complicated interactions at the vertex. These non--trivial vertex functions actually do appear in a renormalization group flow via the $n-1$ Pachner moves of state sum models, which is why using the space of all models of the form (\ref{Eq:VertexFunctionHolonomyStateSum}) allows for computing an exact renormalization group flow in this case. The cost of this is that this space is much larger than the space of $E$-function models, using state sums of the form (\ref{Eq:HolonomyStateSum}). Still, the example in section \ref{p41}  shows that truncating the flow to this subspace can still capture essential features of the phase space, as well as its fixed point structure.\\[5pt]

The form (\ref{Eq:VertexFunctionHolonomyStateSum}) can be simplified by absorbing the $E$-functions into the vertex functions $W_v$. To this end, we write the $E$-functions as convolution of two functions, i.e.

\begin{eqnarray}
E(g)\;=\;\int dk\;F(k)\,F(gk).
\end{eqnarray}

\noindent The function $F$ has the same symmetries than $E$, i.e. $F(g)=F(g'g(g')^{-1})=F(gh)$ for $g,g'\in g$, $h\in G'$.\footnote{It should be noted that $F$ is not unique, but depends on choosing a collection of phase functions. If, however, $E$ is positive, then $F$ can be chosen positive as well.} Thus we get one more integration over $G$ for each pair $ef$, resulting in
\begin{eqnarray}
Z_{\rm Vertex}\;=\;\int dg_v^f\,dg_{ef}\,dk_{ef}\,\prod_f \omega(G_f)\prod_{ef}F(g_{ef}k_{ef})F(k_{ef})\prod_v W_v(g_v^f,\ldots)
\end{eqnarray}

\begin{figure}[hbt]
\begin{center}
	\psfrag{v}{$v$}
	\psfrag{vd}{$v'$}
	\psfrag{e}{$e$}
	\psfrag{ed}{$e'$}
	\psfrag{f}{$f$}
    \psfrag{hvf}{$h_{v}^f$}
\includegraphics[scale=0.4]{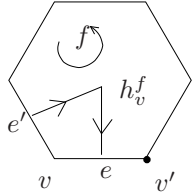}
\caption{\label{Fig:VariableHvf}  Wedge holonomies in the alternative vertex amplitude description.}
\end{center}
\end{figure}

\noindent Performing the variable transformation $g_{ef}\to g_{ef}k_{ef}$ and $g_v^f\to h_v^f\equiv k_{ef}g_v^fg_{e'f}^{-1}$ and carrying out the integration over the $g_{ef}$ and $k_{ef}$, one arrives at the form
\begin{eqnarray}\label{Eq:VertexFunctionHolonomyStateSumAlternative}
Z_{\rm Vertex}\;=\;\int dh_v^f\;\prod_f \omega(G_f)\prod_v V_v\big(h_v^{f},\ldots\big)
\end{eqnarray}

\noindent where $G_f$ is the ordered product of the $k_v^f$ for all $v\subset f$, according to the orientation of $f$. The new vertex functions $V_v$ are given by
\begin{eqnarray}
V_v\big(h_v^{f},\ldots)\;=\;\int dg_{ef}\,\left(\prod_{v\subset e\subset f}F(g_{ef})\right)\;W_v\big(g_{ef} h_v^f g_{e'f}^{-1},\ldots\big)
\end{eqnarray}

\noindent The vertex functions $V_v$ inherit properties from both $W_v$ and $F$. Firstly, it can be readily seen that $V_v$ can -- just as $W_v$ -- be interpreted as a $G$-spin network function on the boundary graph around the vertex $v$. Furthermore, the right-invariance of $F$ under the subgroup $G'$ implies that $V_v$ is invariant under the symmetry
\begin{eqnarray}
h_{v}^f\;\to\;h_1\,h_v^f\,h_2,\qquad h_1,h_2\in G'.
\end{eqnarray}
\noindent Hence $V_v$  is not only a function on elements of $G$, but rather on $\tilde G= G'\backslash G / G'$. If $G'$ happens to be a normal subgroup of $G$, $\tilde G$ in itself is again a group that carries a right and left $G$ action. In this case $V_v$ can be interpreted as function on a graph, depending on $\tilde G$-elements on the edges, invariant under gauge transformations at the vertices with elements from $G$.

It should be noted that one can arrive at the same version of the state sum (\ref{Eq:VertexFunctionHolonomyStateSumAlternative}) by starting from a version of $C_e$-functions and vertex functions $W_v$. Since the $C_e$-functions are more general than the $E$-functions, in this case the resulting vertex functions $V_v$ are just $G$-spin network functions, without any specific properties regarding the subgroup $G'$. It is this form, i.e.~equation (\ref{Eq:VertexFunctionHolonomyStateSumAlternative}) with general $G$-spin network functions $V_v$ and class functions $\omega$, which is the most general of all the presented formalisms.

\begin{table}[h]
\centering
\begin{tabular}{|l|c|c|c|c|}
\hline
State sum & symbol & variables & parameterizing functions \\ \hline \hline
$E$-function & $Z$ & $g_{ev}$, $g_{ef}$ & $\omega$, $E$ \\ \hline
OSFM & $Z_\text{OSFM}$ & $h_{ef}$ & $\omega$, $C_e$ \\ \hline
GFT & $Z_\text{GFT}$ & $g_{vf}^e$ & $\Pi_e$, $\mathcal{V}_v$ \\ \hline
Vertex function & $Z_{\text{Vertex}}$ & $g_v^f$, $g_{ef}$ & $\omega$, $E$, $W_v$ \\ \hline
Vertex function (alternative) & $Z_{\text{Vertex}}$ & $h_v^f$ & $\omega$,  $V_v$ \\ \hline
\end{tabular}
\caption{Different formulations of the Spin Foam state sum, using different variables and parameterizing functions.}
\label{Tab:DifferentFormulations}
\end{table}

\section{The BF, BC and EPRL  models}

Here we will shortly describe the structure of the current spin foam models  \cite{Barrett:1998gs,Engle:2007wy,Freidel:2007py} in the holonomy language.
%Having in hand the relationship to the operator formalism, we can now easily give the formulation of the various established spin foam models in the holonomy language.
%
To do so we specify now to $G = \Spin(4)$ and $G' = \SU(2) = \Spin(4)_{diag}$ the diagonal $\SU(2)$ subgroup, with irreps labelled by $\rho$ and $k$ respectively. Also, the face weights are usually set to be the delta function on the group. % (although this choice leads usually to divergencies).

The operators $\tilde{E}$  described in section \ref{osfm}  for the various models are given by the following, see for instance \cite{alexreview}.

\begin{itemize}
\item
For BF theory we have
\be
{\tilde{E}^{BF}} = \id,
\ee
which reduces the edge projector $P_e$ to be the Haar projector.

\item
In the Barrett Crane model the maps $\tilde E_\rho$ which act on representation space $\rho$, restrict to representations $\rho=(k,k)$. Furthermore  $\tilde E_\rho$  just involves the injection maps (described in section \ref{par})  $I(\rho,0)$ from $\bar\rho$   to the representation $k=0$:
\be
{\tilde{E}^{BC}_\rho} = \dd_e(\rho) \sum_{k} \delta(\rho,(k,k)) I(\rho,0) I(\rho,0)^\dagger \q.
\ee
where $\dd_e(\rho)$ is an unspecified edge measure factor.

\item
For the EPRL model the maps $\tilde E_\rho$ also restrict to a $k$--parameter family of representations, which is now of the form $\rho_\gamma(k) = \left(\frac{1+\gamma}{2} k, \frac{|1-\gamma|}{2}k\right)$ with $\gamma$ being the Barbero--Immirzi parameter:
\be
{\tilde{E}^{EPRL}}_\rho = \dd_e(\rho) \sum_{k} \delta\left(\rho,\rho_\gamma(k)\right) I(\rho, k) {I(\rho,k)}^\dagger \q .
\ee
The injection map picks out a representation $k_\rho$ which is determined by $\rho$ via the relation $\rho=\rho_\gamma(k_\rho)$.  

\item
The FK model is constructed using coherent states \cite{etsim}, a description of the corresponding edge operators and functions can be found in  \cite{alexreview,HS}.

\end{itemize}

The edge functions $E(h)$ can be obtained from the edge operators $\tilde E$  by Fourier transform and using the parameterization in section \ref{par}.
\begin{itemize}
\item
For the BF model we just have $E^{BF}(h)=\delta_G(h)$, i.e. the $E$ function is given by the delta function on the group.
\item
For the Barrett Crane model we obtain
%\be
%E^{BC}(h) = \sum_{\rho, k, k'} \dim(\rho) \dd_e(\rho) \delta(k) \delta\left(\rho,(k',k')\right) \tr_\rho\left(D_\rho(h)\, I(\rho,k) I(\rho,k)^\dagger\right).
%\ee
\be
E^{BC}(h) = \sum_{\rho, k}  \dd_e(\rho) \,\,\dim(\rho)\,\, \delta\left(\rho,(k,k)\right) \tr_\rho\left(D_\rho(h)\, I(\rho,0) I(\rho,0)^\dagger\right).
\ee
For the choice $\dd_e(\rho) = 1$ this results into
\be
E^{BC}(h) = \delta_{SU(2)}(h^+{h^-}^{-1}) = \Theta_{\Spin(4)_{diag}}(h),
\ee

where, for a general subgroup $G' \subset G$ we write \be\Theta_{G'}(h) = \int_{G'} \dd h' \delta(hh'^{-1})\ee for the delta function that forces a group element to lie in the subgroup. Note that the $E$--function has an enhanced symmetry with respect to the subgroup $G'=\Spin(4)_{diag}$, it is not only invariant under conjugation with elements in $G'$ but also invariant under left and right multiplications. This explains some of the special properties of the BC model as compared to more general models.

%Thus we arrive at a particularly simple form for the Barrett-Crane model as an integral over a product of $\SU(2)$ delta functions:

%\be\label{eq-PartFuncBC-OnlyDeltas}
%\ZZ_{BC} = \int \left(\prod_{e \subset f}  \dd h^\pm_{ef}\right) \left( \prod_{v \subset e} \dd g^\pm_{ev} \right) \left(\prod_{e \subset f} \delta\left(h^+ \left(h^-\right)^{-1}\right)\right) \left(\prod_f \delta(g^+_f)\delta(g^-_f)\right).
%\ee

\item
 For the EPRL model the function $E^{EPRL}(h)$ is given by
\be
E^{EPRL}(h) = \sum_{\rho, k} d_e(\rho) \,\, \dim(\rho)\,\, \delta\left(\rho,\rho_\gamma(k)\right) \,\,\tr_\rho\left(D_\rho(h)\, I(\rho,k) I(\rho,k)^\dagger\right) \q .
\ee
As for the BC model this will be a distribution, with singular support described below.

\end{itemize}

The $E$ functions are distributional for all the cases described here and  have support on the following critical manifolds: \begin{itemize}
 \item BF: $\bf{1}$
 \item BC: $\SU(2)_{diag}$
 \item EPRL: $M_{\gamma} = \{g \in Spin(4) | \exists L \in \su(2) : g=(\exp(L), \exp(\gamma L))\}$ \q .
\end{itemize}

For the EPRL model this is shown in reference \cite{SL}. For our purposes it will be convenient to characterise the EPRL support in terms of the group. This can be done for example by choosing a $\U(1)$ subgroup of $\U(1)_{\gamma} \in \Spin(4)$ generated by $(L, \gamma L)$. The critical manifold is then simply $\SU(2)_{diag} \triangleright_{ad} \U(1)_{\gamma}$, where $\triangleright_{ad}$ is the adjoint action.

\subsection{General (finite) groups}

We can now mimick the three types of spin foam models defined above for finite groups. To do so we assume that the most important part of the amplitude is the divergent, distributional part.
To define models for finite groups we choose functions with support on the appropriate subsets of $M \subset G$, that is by replacing the delta functions by normalized $\Theta_{M} (g)$ functions that are defined to be $\frac1{|M|} \forall g \in M$ and $0$ otherwise. The $E$-functions  are then:

\begin{itemize}
 \item BF: $E(h) = \Theta_{e}(h)$ with $e$ being the trivial subgroup consisting of the identity element.
 \item BC: $E(h) = \Theta_{G'}(h)$
 \item EPRL: $E_{C}(h) = \Theta_{M_C}(h)$, where $C$ is a cyclic subgroup of $G$ and $M_C = G' \triangleright_{ad} C$.
\end{itemize}

This generalizes the BC and the EPRL model to arbitrary groups $G' \subset G$. %They are elements of the finite dimensional linear space $\EE(G, G')$.

\subsubsection{Example: $S_3$}\label{s3ex}

 Let us consider as an example  the finite group $G = S_3$, the group of permutations of three elements, as gauge group. It can be written as
\begin{eqnarray}
S_3\;=\;\langle a,b|a^2=b^3=(ab)^2=1\rangle.
\end{eqnarray}

The group has six elements $\{e, a, bab^{-1}, b^2ab^{-2}, b,b^2\}$, where $e$ is the unit element, $\{a,bab^{-1},b^2ab^{-2}\}$ are odd permutations (two--cycles) and $\{b,b^2\}$ are three--cycles.

The group has subgroups isomorphic to $\mathbb{Z}_2$,  generated by either $a$, $ab$ and $ab^2$. Furthermore there is a $\mathbb{Z}_3$ subgroup ${e,b,b^2}$ which is normal in $S_3$.
To specify the models further we have to choose one of the subgroups and consider $E$ functions which are invariant under the adjoint action of the subgroup $G'$ on $G$. The orbits under the different subgroups, including the group $G=S_3$ are
\begin{eqnarray}\label{orbits}
G'\equiv S_3:&&\{e\},\{ a, bab^{-1}, b^2ab^{-2}\},\{b,b^2\}; \nn\\
G'=\{e,a\}\equiv\mathbb{Z}_2:&& \{e\},\{a\},\{bab^{-1}, b^2ab^{-2}\}, \{b,b^2\}; \nn\\
G'=\{e,b,b^2\}\equiv{\mathbb Z}_3: && \{e\},\{ a, bab^{-1}, b^2ab^{-2}\}, \{b\},\{b^2\} \q .
\end{eqnarray}

Choosing $G'=\mathbb{Z}_2$ as subgroup, we have a priori a four parameter space of $E$--functions
\bea\label{gpar}
E(g)=\beta_1 \delta(g,e) + \beta_2 \delta(g,a) + \beta_3(\delta(g,bab^{-1})  +\delta(g,b^2ab^{-2})) + \beta_4 (\delta(g,b)+\delta(g,b^2)). \nn\\
\eea
 Taking some normalization condition into account, we obtain three parameters.
 As will be explained in section \ref{cedges}, models in which the $E$ function is invariant under conjugation of the full group can be rewritten as standard lattice gauge theory models, i.e. the $E$ functions can be absorbed into the face weights $\omega$. This will be the case for $\beta_2=\beta_3$.

The BC model for the choice of this subgroup $\mathbb{Z}_2$ corresponds to $\beta_1=\beta_2=1$ and $\beta_3=\beta_4=0$. Hence it will in general define a non--trivial model, which cannot be rewritten into a lattice gauge theory. %{\red As we will discuss in section \ref{}, even in the three--valent case the model is not factorizable, i.e. rewritable into a gauge theory. }

We can define an EPRL like model with $\mathbb{Z}_3$ as cyclic subgroup, which leads to $\beta_1=\beta_4=1$ and $\beta_2=\beta_3=0$. This model is also a $\mathbb{Z}_3$ subgroup model (the BC model with subgroup $\mathbb{Z}_3$) and actually rewritable as a standard lattice gauge theory. For this choice of parameters we have
\bea
E(g)=\Theta_{\mathbb{Z}_3}(g)
\eea
and as $\mathbb{Z}_3$ is a normal subgroup the function $E$ will be invariant under conjugation. Indeed the $BC$ model with a normal subgroup will define a topological theory which is of $BF$ type on the quotient group (here $\mathbb{Z}_2$) and of high temperature type (i.e. $E$ is a constant function) on the subgroup itself.  We can define two different one-parameter families which connect this topological theory to either the high temperature/strong coupling fixed point of the full group ($\beta_1=\beta_2=\beta_3=\beta_4$)
\bea
E(g)=\Theta_{\mathbb{Z}_3}(g)+\gamma \,\Theta_{odd}(g)
\eea
(here $\Theta_{odd}=1$ on odd permutations and vanishes on even permutations)
or to the BF theory on the full group
\bea
E(g)=\delta(g,e) + \beta_4 (\delta(g,b)+\delta(g,b^2)) \q .
\eea

%The EPRL model with $\mathbb{Z}_2$ as cyclic subgroup and $G'=\mathbb{Z}_3$ would have $\beta_1=\beta_2=\beta_3=1$ and $\beta_4=0$ and therefore would again be equivalent to some lattice gauge theory.

Another example to define a BC model more closely connected to the $SU(2)\times SU(2)$ model is to take $G=S_3\times S_3$ as gauge group and the diagonal group as subgroup. We will discuss this model shortly in section \ref{c3d}. % We can also define an EPRL model in $S_3\times S_3$ where the cyclic group is given by $\{(1,1),(b,b^2),(b^2,b)\}$. This group is already invariant under the adjoint action of the diagonal subgroup in $S_3\times S_3$, hence the corresponding $E$ functions has support just on this cylic group and coincides with the  $BC$ model.

\section{Coarse graining}\label{coarse}

In this section we will consider coarse graining of the holonomy spin foam
models in dimensions two and three. To this end we will utilize in the first part techniques from standard lattice gauge theory, that is use the similarity of holonomy spin foams to standard gauge theories. Indeed, as for gauge theories, coarse graining in two dimensions can be done exactly. The holonomy spin foam models feature in addition to the face weights, which are the central dynamical entitles for standard lattice gauge theory, the edge weights or edge functions $E$, which will lead to additional factors in the coarse graining formulas.

In three dimensions certain models, including BC type models, can be rewritten into models of standard lattice gauge theory type. We will identify this class of models and coarse grain these models by applying the Migdal Kadanoff scheme.

These results can be compared to another type of coarse graining scheme, which involves a hierarchical lattice on which coarse graining can be performed by applying repeatedly $4-1$ Pachner moves. This allows to do an exact coarse graining. In particular we will test whether the parametrization of the holonomy models with $E$ function captures the relevant dynamical data. That is although the general coarse graining flow leaves the model space described by the $E$ functions we can project back to this space and test whether we still find the fixed point structure of the exact flow.

Attempting concrete (numerical) calculations we face the difficulty that the models are not necessarily finite for (compact) Lie groups.
To avoid these difficulties numerical investigations will be (mostly) performed with finite groups, in particular $S_3$, for which we  defined a class of models in the last section.

\subsection{Coarse graining of edges}\label{cedges}

Let us consider the formulation of the models (\ref{Eq:OSFMCFunctions}), where to every edge we associate a $C$--function and to every face face weights $\omega$. One important difference of these models to standard lattice gauge theories is, that in general, these are not invariant under edge subdivision, or the inverse operation, coarse graining of edges by removing two--valent vertices.

Therefore we will briefly discuss the coarse graining of edges. This is most conveniently done in the spin basis
where the $C$--function is expressed as follows
\bea\label{d06gen1}
C(h_1,h_2,\cdots h_n)=\!\!\!\!\!\sum_{\rho_1,\cdots, \rho_n, dd'}
\prod_{i=1,\ldots,n}\dim\rho_i\,
\tilde C_{\rho_1,\cdots,\rho_n,dd'}
\left\langle\iota_{d'}\,|\,
D_{\rho_1}(h_1)\otimes\cdots\otimes
D_{\rho_n}(h_n)\iota_d\right\rangle.\nn\\
\quad
\eea
Here  $\{\iota_d\}$ is an orthonormal basis of the intertwiners and $\left\langle \cdot \,|\,\cdot\right\rangle$ is the inner product on $L^2(G^n)$.

The coefficients $\tilde{C}$ can be computed from the coefficients $e^k_{\rho dd'}$ defined in (\ref{Eq:DecompositionOfEFunctions}) if the $C$--function is derived from an $E$--function as in
\begin{equation}
 C(h_{ef_1},\ldots, h_{ef_n})=\int \dd g_L\dd g_R \prod_i \dd \tilde{h}_{ef_i}
\prod_i E(\tilde{h}_{ef_i})\prod_i \delta(g_Lh_{ef_i} g_R \tilde{h}_{ef_i}^{-1}) \q .
\end{equation}
Expanding the $\delta$ function $\delta(g)=\sum_\rho \dim\rho\, \chi_{\rho}(g)$ and
using the identity
\bea
 \int \dd g_L \dd g_R\, &&\prod_i \chi_{\rho_i}(g_Lh_{ef_i} g_R
\tilde{h}_{ef_i}^{-1})
= \nn\\
&&\sum_{d\, d'}{\rm tr}\, |\iota_{d'}\rangle\langle\iota_{d'}|
\bigotimes D_{\rho_i}(h_{ef_i})
|\iota_d\rangle\langle\iota_d|\bigotimes D_{\rho_i}(\tilde{h}_{ef_i}^{-1})
\eea
we obtain
\begin{equation}
 \tilde{C}_{\rho_1\ldots,\rho_n\,dd'}=
\prod_i \dd \tilde{h}_{ef_i}
\prod_i E(\tilde{h}_{ef_i})
\left\langle\iota_d| \bigotimes
D_{\rho_i}(\tilde{h}_{ef_i}^{-1})\iota_{d'}\right\rangle \q .
\end{equation}
Using the explicit form  for the $E$ function (\ref{Eq:DecompositionOfEFunctions}) yields
\begin{equation}\label{d06gen4}
 \tilde{C}_{\rho_1\ldots,\rho_n\,dd'}=
\sum_{k_i, d_id_i'}\prod_i e^{\rho_i}_{k_id_id_i'}\,
\langle\iota_d|\bigotimes_i
I(\rho_i,k_i)_{d_i}I^\dagger(\rho_i,k_i)_{d_i'}|\iota_{d'}\rangle \q .
\end{equation}

\begin{figure}[hbt]
\begin{center}
	\psfrag{e}{}
    \psfrag{e1}{$e$}
    \psfrag{e2}{$e'$}
    \psfrag{f}{$f$}
    \psfrag{fdash}{$f'$}
    \psfrag{R}{$\longrightarrow$}
\includegraphics[scale=0.4]{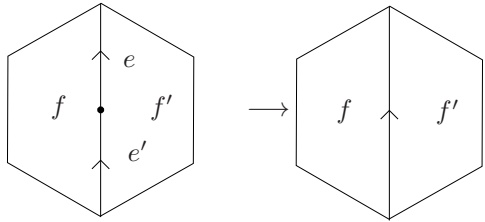}
\caption{\label{edge sub}Removing a trivial subdivision of an edge.}
\end{center}
\end{figure}

With the form \eqref{d06gen1} of our edge function $C$ at hand we can now
consider the coarse graining, where a subdivided edge, i.e. two edges $e,e'$
joint by a two--valent vertex, is coarse grained into one edge $n$. All edges
bound the same faces $f_1,\ldots,f_m$, see Figure \ref{edge sub}. As $C$--functions
are composed by delta--functions, the $C$--function for the new edge $n$ is given
by
\begin{equation}
\label{d06gen2}
\begin{split}
C^{(n)}(h_{nf_1},&\ldots,h_{nf_m})
=
\int \prod_i\dd h_{ef_i} \prod_i \dd h_{e'f_i}\\
&C^{(e)}(h_{ef_1},\ldots, h_{ef_m})\,C^{(e')}(h_{e'f_1},\ldots,
h_{e'f_m}) \prod_i\delta(h_{nf_i}^{-1} h_{ef_i}h_{e'f_i})  \q .
 \end{split}
\end{equation}
Expanding all functions involved as in \eqref{d06gen1} and using the orthogonality
of the basis $\iota_d$ we arrive at
\begin{equation}\label{d06gen3}
 \tilde C^{(n)}_{\rho_1,\cdots,\rho_n,dd'}=
\sum_{d''}\tilde C^{(e)}_{\rho_1,\cdots,\rho_n,dd''}\tilde
C^{(e')}_{\rho_1,\cdots,\rho_n,d''d'}\quad .
\end{equation}
This gives the renormalization flow equation for the coarse graining of edges by removing two--valent vertices. Note that the flow equations for this operation are local.

In 2D and many 3D models the $C$--function simplifies drastically, so that the models can be reduced to standard gauge theory.  To explain how this happens,
we define an $n$-valent $C$--function to be factorizable, if it can be written as
\begin{equation}
 \label{eq-factor}
C(h_{ef_1},h_{ef_2},\ldots, h_{ef_n})=\int \dd g_L  \, \epsilon(g_L h_{ef_1})\,\,
\epsilon(g_L h_{ef_2})
 \cdots \epsilon(g_L h_{ef_n})
\quad .
\end{equation}
where $\epsilon$ is a class function. Let us notice that every function of the form
\eqref{eq-factor} can be rewritten into a left and right group averaging of $E$--functions since due to the symmetries of
$\epsilon$ we have
\begin{equation}
C(h_{ef_1},h_{ef_2},\ldots, h_{ef_n})=\int \dd g_L\dd g_R  \, \epsilon(g_L h_{ef_1}g_R)\,\,
\epsilon(g_L h_{ef_2}g_R)
 \cdots \epsilon(g_L h_{ef_n}g_R)
\quad ,
\end{equation}
hence $E=\epsilon$.
The form introduced above allows us to integrate out $h_{ef}$ and to absorb the $\epsilon$ factors into the face weights $\omega$. If we define the effective face weights
\begin{equation}
 \omega'_f=\int  \prod_{e \subset f} \dd h_{ef} \epsilon(h_{ef})
\, \omega_F (\cdots g_{ve}h_{ef}g_{ev'}\cdots) \q ,
\end{equation}
these are again class functions of the group element $\cdots g_{ve}g_{ev'}\cdots$. (Note that the face weights $\omega'_f$ might depend on the face $f$, in particular if faces differ in the number of adjacent edges.) Explicitly, using the Fourier transform
\begin{align}
 {\omega}'(g)&=\sum_\rho \dim\rho \, {\omega'}^\rho \chi_\rho(g)\\
\epsilon(g)&=\sum_\rho \dim\rho \,\epsilon^\rho \chi_\rho(g)
\end{align}
we obtain
\begin{equation}
 {\tilde\omega'}{}\!\!^\rho \,=\, (\epsilon^\rho)^{\sharp(e\subset f)} \,\, \tilde{\omega}^\rho .
\end{equation}
In this way every spin foam with factorizable $C$--functions can be rewritten as a standard lattice
gauge theory. This factorization property holds for every two--valent edge and in the
case of BC/EPRL/FK also for every three--valent edge.

If we do not integrate out the $h_{ef}$ and keep the $C$--functions, the Fourier transform of these will be given by
\begin{equation}\label{w2}
 \tilde C_{\rho_1,\cdots,\rho_n, dd'} =\left(\prod_i \epsilon^{\rho_i}\right)\,
\delta_{dd'} \q .
\end{equation}
This will simplify the renormalization flow (\ref{d06gen3}).

\subsection{Coarse graining in two dimensions}

Here we will consider a 2--complex for which all edges are 2-valent, i.e. bound exactly two faces. In this case the coarse graining flow can be computed exactly. The procedure  and result is quite similar to real space renormalization of 2D lattice gauge theories \cite{Migdal:1975zg}. As pointed out in section \ref{cedges}, compared to standard lattice gauge theory one obtains however an additional renormalization flow from the edge weights.
Starting with the form of the partition function
\bea\label{2d01}
Z%&=& \int \left(\prod_{(ve)} \dd g_{ve}\right)     \left(\prod_{(fe)} \dd h_{fe}\right)   \left(\prod_{(fe)} E(h_{ef})\right)    \left( \prod_f \omega_f(g_{ve}h_{ef} g_{ev'} g_{v'e'} h_{e'f}\cdots)\right) \nn\\
&=&\int    \left(\prod_{(fe)} \dd h_{fe}\right)   \left(\prod_{(e)} C(h_{ef},h_{ef'})\right)    \left( \prod_f \omega_f(h_{ef} h_{e'f}\cdots)\right) \quad ,
\eea
we have to discuss the renormalization flow of the $C$-functions and the face weights $\omega_f$

%Here the edge functions are given as
%\be\label{2d02}
%C(h_{ef},h_{ef'})=\int \dd g_L \dd g_R E(g_L h_{ef} g_R) \, E(g_L h_{ef'} g_R) \, ,
%\ee
%that is $C$ is a left and right average of a square of functions.

The $C$--function is by construction invariant under left and right multiplication (in both entries). Hence it can be expanded into a basis of gauge invariant functions, that is
\be\label{2d03}
C(h_1,h_2)=\sum_\rho   \tilde C_\rho  \text{dim}(\rho) \chi_\rho(h_1 h_2^{-1}) \quad .
\ee
%Here $\chi_\rho(\cdot)=\tr_\rho(D_\rho(\cdot))$ is the character of the representation $\rho$.

Such a $C$--function is factorable in the sense of  (\ref{eq-factor}): the Fourier expansion of the factorization property gives
\be
C(h_1,h_2)=\int \dd g \epsilon(gh_1) \epsilon(gh_2)\;=\; \sum_\rho  \epsilon^\rho \epsilon^{\bar{\rho}} \, \text{dim}(\rho) \chi_\rho(h_1 h_2^{-1})  \q .
\ee
Hence we just need to choose the Fourier coefficients $\epsilon^\rho$ such that $\epsilon^\rho\epsilon^{\bar\rho}=\tilde C_\rho$ is satisfied.

The coefficients  $\tilde C_\rho$ can be expressed using
the parametrisations through $e^\rho_{k,dd'}$ as \begin{equation}\label{2d04rep}
 \tilde{C}_\rho=\sum_{k} \frac{\dim k}{\dim\rho} \sum_{dd'} e^\rho_{kdd'}e^\rho_{kd'd}
\end{equation}
where we used that $e^{\bar{\rho}}_{\bar{k}dd'}=e^{\rho}_{kd'd}$, which can be shown to hold in general \cite{HS}.
That is from the choices $e^\rho_{kdd'}$ for the basis coefficients of $E$ only the
contraction (\ref{2d04rep}) matters in a 2--dimensional theory.

%With the form (\ref{2d03}) of our edge function $C$ at hand we can now consider
%the coarse graining, where a subdivided edge, i.e. two edges $e,e'$
%joint by a two--valent vertex, is coarse grained into one edge $n$. All edges
%bound the same two faces $f,f'$.

The renormalization flow equations for the coarse graining of edges (\ref{d06gen3}) are in this case given by
\bea\label{2d07}
\tilde C^{(n)}_\rho &=&\tilde C^{(e)}_\rho \tilde C^{(e')}_\rho  \quad .
\eea

\begin{figure}[hbt]
\begin{center}
	\psfrag{e}{$e$}
    \psfrag{f}{$f$}
    \psfrag{fdash}{$f'$}
    \psfrag{fddash}{$f''$}
    \psfrag{R}{$\longrightarrow$}
\includegraphics[scale=0.4]{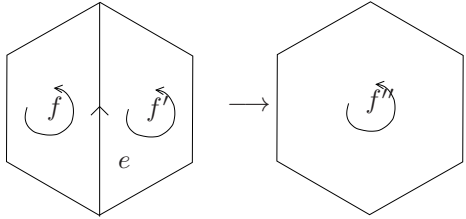}
\caption{\label{facesub} Removing a  subdivision of a face.}
\end{center}
\end{figure}

The other coarse graining step that will appear in two dimensions is integrating out the variables over one edge  $e$ that divides two faces $f,f'$, see Figure \ref{facesub}. The face holonomies starting at the source of $e$ will be denoted by $g_f=h_{ef}H_f$ and $g_{f'}=h_{ef'}H_{f'}$. The new face weight is given by
\bea\label{2d08}
\omega_{f\cup f}^{(n)}(H_f H_{f'}^{-1})&=& \int \dd h_{ef} \dd h_{ef'} \,\omega_f(h_{ef}H_f) \, \omega_{f'}(h_{ef'}H_{f'}) C(h_{ef},h_{ef'}) \quad ,
\eea
%We again Fourier expand all factors in the integrand of (\ref{2d08}), where for the face weights we use the convention $\omega=\sum_\rho \tilde \omega(\rho) \text{dim}(\rho) \chi_\rho$. This results in the new face weight coefficients
which for the Fourier coefficients gives
\bea\label{2d09}
\tilde \omega^{(n)}_{f\cup f}&=& \tilde \omega_f(\rho) \, \tilde \omega_{f'}(\rho) \,\, \tilde C_\rho ^{(e)}\quad .
\eea

In two dimensions all coarse graining steps can be decomposed into the two operations we discussed, that is the inverse of subdividing a face with one edge and the inverse of subdividing an edge with one vertex. This allows us to write the complete partition function of a 2--dimensional complex as
\bea
Z=\sum_\rho (\tilde \omega(\rho))^F ( \tilde C_\rho)^E
\eea
(assuming that the face weights and edge functions are initially the same for all faces and edges).

The recursion relations (\ref{2d07}) and (\ref{2d09}) enable us to discuss completely the renormalization flow in two dimensions. Starting with a 'heat kernel' ($\cal{C}_\rho$ is the Casimir of $\rho$)
\bea
\tilde C_\rho=\exp( -\beta^{-1}\cal{C}_\rho)
\eea
for the edge weights, we flow from zero temperature $\beta=\infty$ to infinite temperature $\beta=0$. Note that zero temperature $\beta=\infty, \tilde C_\rho=1$ corresponds to a $C$--function $C(h_1,h_2)=\delta(h_1h_2^{-1})$. This converts the partition function (\ref{2d01}) into  one for standard lattice gauge theory.  Choosing non-zero temperature corresponds to a regularization (introducing a non--trivial propagator for the edges) for instance in group field theory \cite{Geloun:2010vj}. One should however be aware that this choice leads to a flow into the high temperature fixed point, both for the edge factors and the face weights. In this fixed point only the values for the trivial representation contribute (assuming ${\cal{C}}_{\rho=0}=0$).

The renormalization flow for the face weights behaves similarly, with the addition that it can be set off from the zero temperature fixed point $\omega(g)=\delta(g)$ by a non-trivial choice for the edge factors alone. Note that the zero and high temperature fixed points are not the only fixed points, as will be obvious from the discussion below.

In higher dimensions the  subdivision of an edge by adding a two--valent vertex, as well as the  subdivision of a face by inserting a two--valent edge can be argued to be operations under which the partition function should remain invariant \cite{Ben}. Indeed this notion has been connected to the anomaly--freeness of the path integral measure in spin foams with respect to diffeomorphisms \cite{Bojowald:2009im,Bahr:2009qc,Bahr:2011uj} and has the advantage of fixing uniquely the edge and face measure factors \cite{Ben}. Such an invariance would require $\tilde C_\rho$ and $\tilde \omega(\rho)$ to take values either equal to one or to zero.  Any choice would lead to a fixed point in the two--dimensional theory. Also note that for the two--dimensional theory it is equivalent to either choose $\omega(\rho)=0$ or $\tilde C_\rho=0$ for a given $\rho$. That is all $\tilde C_\rho$ can be chosen to be equal to one. Hence we just need to consider the lattice gauge theory parameter space, reduced to the choice $\tilde \omega(\rho)=\theta(\rho)$ where $\theta$ takes only values $0$ and $1$.

\subsection{Dual simplicial lattices in three dimensions}\label{c3d}

We have discussed that the renormalization flow for the two--dimensional theory agrees with the one in lattice gauge theory in case that one adopts the condition of invariance under edge subdivisions. This conditions is also a reasonable requirement to hold for higher dimensions and higher-valent edges. Indeed the Barrett-Crane model can be easily made to satisfy this condition. For the EPRL model the construction \cite{Kaminski:2009fm,Bahr:2010bs} is more complicated and leads to a model that falls not anymore into the description by $E$-functions (would however be still in the more general class of models parametrized by $C$-functions).

\begin{figure}[hbt]
\begin{center}
	\psfrag{h1}{$h_1$}
	\psfrag{h2}{$h_2$}
	\psfrag{h3}{$h_3$}
	\psfrag{r1}{$\rho_1$}
	\psfrag{r2}{$\rho_2$}
	\psfrag{r3}{$\rho_3$}
    \psfrag{i}{$\iota$}
    \psfrag{id}{$\iota^{\dag}$}
    \psfrag{=}{  
    %$=\;\;\;\;\psi_{\rho_1,\rho_2,\rho_3,\iota}(h_1,h_2,h_2)$
  }
\includegraphics[scale=0.5]{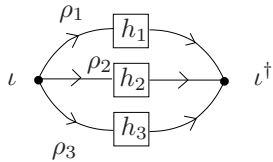}
\caption{\label{3i}  The structure of the $C$ function for an edge shared by three faces.  }
\end{center}
\end{figure}

Let us assume that we can construct a spin foam model where such an invariance
holds. How much freedom is left for the parameters in this spin foam model? In a
dual simplicial lattice in three dimensions the edges (dual to triangles) would
be three---valent, i.e. adjacent to always three faces (as triangle has three
edges). The $C$--function associated to such three--valent edges is a function
of three arguments, invariant under left and right group multiplication. Hence
it can be expanded into a basis of gauge invariant spin networks, which are
based on three edges with common source and target vertex, see Figure \ref{3i}.

These spin networks are labelled with representations at the edges and with
intertwiners between the three representations, that is we have
\bea\label{3d01}
C(h_1,h_2,h_3)=\sum_{\rho_1,\rho_2,\rho_3, dd'}
\prod_i\dim\rho_i\,
\tilde C_{\rho_1,\rho_2,\rho_3,dd'}
\left\langle\iota_{d'},\,
D_{\rho_1}(h_1)\otimes D_{\rho_1}(h_2)\otimes
D_{\rho_n}(h_3)\iota_d\right\rangle
\quad .
\eea
where $\iota_d$ labels orthonormal basis of invariants.
We will consider groups, such as $SU(2),SO(4)$ and the finite group $S_3$ for which the intertwiner
between three representations is canonical, or in other words for which the tensor product of two
representations is multiplicity free. In this case the intertwiner labels $d,d'$
in (\ref{3d01}) can
be omitted. Also the sum in (\ref{3d01}) is only over triples of representations admitting an
intertwiner.

Gluing two edges with a two--valent vertex leads to new $\tilde C$--coefficients
\bea\label{3d02}
\tilde C_{\rho_1,\rho_2,\rho_3}^{(n)}=\tilde C_{\rho_1,\rho_2,\rho_3}^{(e)} \tilde
C_{\rho_1,\rho_2,\rho_3}^{(e')}
\eea

Demanding invariance under edge subdivision fixes the coefficients to be equal to one or zero, depending on the three representation labels. In the BC and EPRL model (with $G$ isomorphic to $G'\times G'$ group) this condition even factorizes as in equation (\ref{w2}), that is we have
\bea\label{3d03}
\tilde C_{\rho_1,\rho_2,\rho_3}=\epsilon^{\rho_1}\epsilon^{\rho_2}\epsilon^{\rho_3}
\eea
where $\epsilon_\rho$ takes only the values zero or one.
%This is the same factorization property as in (\ref{eq-factor},\ref{w2}).
One might expect such a factorization property to hold from the construction of the edge
propagator $C$ as group average of a product of $E$--functions. Note however,
that even in this class of theories the factorization property does not hold in
general. For a counter example see appendix \ref{app-example}.  Also the BC model for $S_3$ with $\mathbb{Z}_2$ as a subgroup provides a counter example. As one can check explicitly, the corresponding three-valent edge function $C(h_1,h_2,h_3)$ does not lie in the subspace generated by the conjugation invariant $E$--functions, which define factorable models.

The following considerations restrict therefore to the class of theories where the coefficients factorize  as in (\ref{3d03}). In this case we can absorb the choice function $\epsilon_{\rho}$ into the face weights $\tilde \omega_f$, whereas the $\tilde C$ can be chosen to be the same as for standard lattice gauge theory (where invariance under edge subdivision trivially holds). In other words we are reduced again to consider the class of theories described by standard lattice gauge theories. This motivates us to test approximation methods developed for lattice gauge theory, notable the Migdal--Kadanoff approximation \cite{Migdal:1975zg,Kadanoff:1976jb}. There is one draw--back which is that the Migdal--Kadanoff approximation was developed for regular (cubical) two--complexes and for hierarchical lattices \cite{Ito} (where it can be made to be exact). That is we have to assume that the difference of a simplicial versus a cubical lattice does not influence the large scale dynamics.

The Migdal--Kadanoff approximation is a truncation to local couplings for the real space renormalization flow of lattice gauge theories. As for the flow in two dimensions it results into a recursion relation for the face weights $\omega_f$. In the version of \cite{Migdal:1975zg} we have for the coarse graining of a three--dimensional cubical lattice under the doubling of the lattice constant
\bea\label{3d04}
\tilde \omega^{(n)}_\rho=\sum_{\rho_1,\rho_2} (\tilde \omega_{\rho_1})^4 (\tilde \omega_{\rho_2})^4 \frac{\text{dim}(\rho_1) \text{dim}(\rho_2)}{\text{dim}(\rho)} n^\rho_{\rho_1,\rho_2} \quad .
\eea
Here $n^\rho_{\rho_1,\rho_2}$ is the number of copies of representations  $\rho$ in the tensor product of $\rho_1$ with $\rho_2$. For the case we are considering here, this number is either one or zero.

Note the essential difference of (\ref{3d04}) to the two--dimensional recursion relation (\ref{2d08}) which is that the coefficients of different representation labels are coupled with each other. In the BC and EPRL spin foam models for a group $G=G'\times G'$ only a (one-parameter) subset of all possible representations is allowed initially, however the recursion relation (\ref{3d04}) will spread the range of allowed values. %This might explain the result, which we found for the BC models, namely that it converges very fast to the high temperature/ strong coupling limit of lattice gauge theory where only the trivial representation has non-zero amplitude.

The implementation of (\ref{3d04}) for a finite group is straightforward. Here we consider the group $S_3\times S_3$ with a BC like amplitude. There are three representations for $S_3$ which we denote by $j=0,1,2$. The representation $j=2$ is two--dimensional whereas the others are one--dimensional, given by the trivial and the (permutation) sign representation. For $S_3\times S_3$ we have representations $(j^+,j^-)$ with $j^+,j^-=0,1,2$. The initial face weights are given by
\bea\label{3d05}
\tilde \omega_{j^+,j^-}=\delta_{j^+,j^-} \quad,
\eea
which we put into the recursion relation (\ref{3d04}). The coupling coefficients are given by
\bea
n^{(j^+,j^-)}_{(j^+_1,j^-_1),(j^+_2,j^-_2)}=n^{j^+}_{j^+_1,j^+_2} n^{j^-}_{j^-_1,j^-_2}
\eea
where $n^{j_3}_{j_1,j_2}$ is invariant under permutations of the three labels and equal to one for the following combinations of representation labels
\bea
(0,0,0),\,\,  (0,1,1),\,\, (0,2,2),\,\, (1,2,2),\,\, (2,2,2) \quad .
\eea
For all other combinations the coefficients are vanishing.
After each renormalization step we normalize the amplitude so that $\tilde \omega_{0,0}=1$. The result is that after ca. 10 iterations  the theory is at a fixed point
\bea
\tilde \omega_{j^+,j^-}=(\delta_{j^+,0}  +\delta_{j^+,1})( \delta_{j^-,0}  + \delta_{j^-,1} )  \quad,
\eea
These weights correspond (modulo normalization) to a face weight
\bea
\omega(g_1,g_2)=\Theta_{even}(g_1)\Theta_{even}(g_2)
\eea
which defines a BF($\mathbb{Z}_2)$ theory, where this $Z_2$ can be understood to be the the quotient group of the diagonal subgroup with respect to the  $\mathbb{Z}_3$ subgroup of even diagonal elements.

Let us also discuss the $SU(2)\times SU(2)$ BC model. Representations are given by $(j^+,j^-)$ where the spins can take integer or half integer non-negative values and the dimension of the representation $(j^+,j^-)$ is given by $(2j^++1)(2j^-+1)$. The coefficients are  $n^{j_3}_{j_1,j_2}$ is equal to one, if the triangle inequalities $j_i+j_j\geq j_k$
are satisfied and if the sum $j_1+j_2+j_3$ is an entire number, otherwise it is zero. The initial face weights are again given by (\ref{3d05}).

\begin{figure}[hbt]
\begin{center}
	
\includegraphics[scale=0.6]{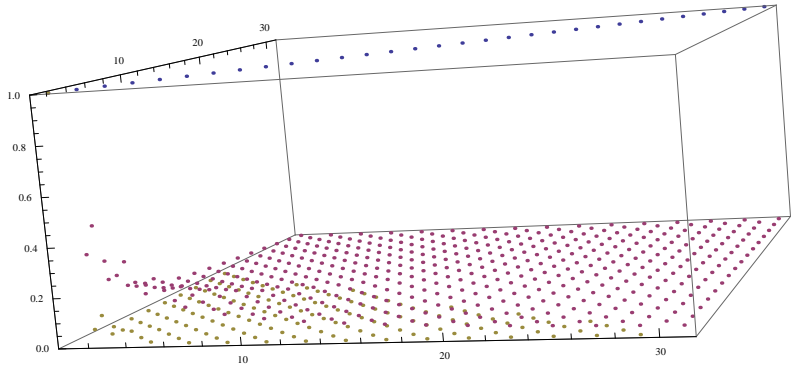}
\label{mkflow}
\end{center}
\caption{ This figure shows the face weights $\tilde \omega_{j^+,j^-}$ as a function of $2j^+,2j^-$, which label the $x$ and $y$ axis respectively. The initial configuration $\tilde \omega_{j^+,j^-}=\delta_{j^+,j^-}$ of face weights is shown in blue. The red dots give the configuration after one iteration, where we show only a dot for values of $\tilde \omega_{j^+,j^-}>10^{-4}$. The cut-off is $\lambda<15$.  As can be seen from the picture the initial diagonal configuration is spread to non--diagonal arguments, but the values for the face weights are decreasing fast. After two iterations  (yellow-green dots) a large portion of the $\tilde \omega_{j^+,j^-}$  is already smaller than $10^{-4}$ and the configuration is almost at the high temperature fixed point.}
\end{figure}

However  the sum (\ref{3d04}) diverges. Therefore we introduce a cut-off and sum values only up to  $j\leq \lambda$.  Again we normalize after each iteration step, such that $\tilde \omega_{0,0}=1$. An impression of the behaviour of the flow can be obtained from figure \ref{mkflow}.

As one can see the face weights flow very fast into the high temperature fixed point and the result does almost not vary with the  cut-off. Note also that whereas initially only a one-parameter subset of face weights is non-zero, this range is spread during the recursion process. Therefore one can also expect a similar flow for the EPRL model, where the initial face weights are also only supported on a one-parameter set of labels.

Of course the introduction of the cut--off can be questioned. On the other hand this result is supported by a general theorem, stating that in 3D gauge theories with groups $U(1), SU(N)$ flow under the Migdal Kadanoff relations to the strong coupling / high temperature fixed point \cite{Ito,schulz}.

\begin{figure}[hbt]
\begin{center}
	\psfrag{r}{$\begin{array}{l}\text{Read off new coefficients by}\\\text{comparing expectation values}\\\text{of boundary functions}\end{array}$}
    \psfrag{p}{$\begin{array}{l}\text{Put new model}\\\text{on $2$-complex}\end{array}$}
    \psfrag{g}{$\begin{array}{l}\text{Glue four}\\\text{$2$-complexes together}\end{array}$}
\includegraphics[scale=0.40]{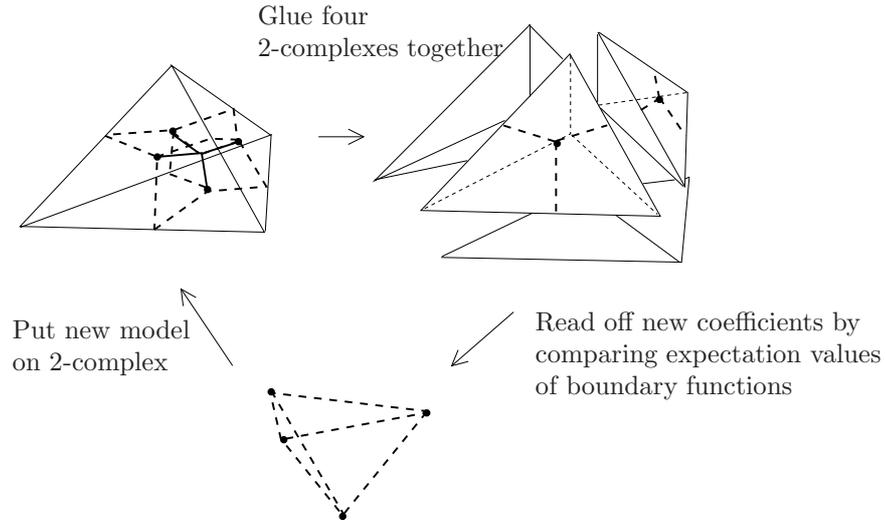}
\caption{The 1-4 Pachner move can be iterated to generate an actual flow in the space of models.}\label{Fig:14move}
\end{center}
\end{figure}

Now, we made several approximations, which could be questioned. The first is to replace a simplicial lattice with a regular one, the other is the Migdal--Kadanoff approximation itself. To support the findings here, we considered an exact renormalization flow for a special simplicial lattice, which will be discussed more extensively in the next section. This lattice is obtained by iteratively applying $1-4$ Pachner moves to a tetrahedron, see figure \ref{Fig:14move}.

Such a $1-4$ Pachner move subdivides one tetrahedron into four. To these four one can again apply the Pachner moves obtaining sixteen, and so on. Going backwards, we can coarse grain blocks of four tetrahedra into one. The advantage of such a hierarchical lattice is that the renormalization flow stays in the set of local couplings. Truncations, which are usually unavoidable in real space renormalization due to the generations of non-local couplings, are not necessary.
This renormalization flow on the simplicial lattice is most convenient to consider in the spin representation with vertex weights. Here one is actually working with the space of vertex weights for an $S_3$ model. In the spin representation the BC model with $S_3\times S_3$ is included in this space. %Then the flow for the $S_3\times S_3$ Barrett Crane model can be found in figure \ref{} \marginpar{\red FIGURE}.
The flow of this model terminates  in a fixed point  describing a BF$(\mathbb{Z}_2)$ theory, which confirms our findings with the Migdal Kadanoff truncation.

 Another example, which we can consider within $G=S_3$ is with weights $\tilde \omega_0=1$ and $\tilde \omega_2=1$.\footnote{This model corresponds to the choice of parameters $\beta_1=5,\beta_2=\beta_3=1,\beta_4=-1$ in (\ref{gpar}) and $e_1=0,e_{20}=e_{21}=1$ in the next section.} With the Migdal Kadanoff recursion relation (\ref{3d04}) it flows to BF theory for the group $S_3$.  With the hierarchical lattice flow, explained in the next section we obtain a different result: here it flows to the strong coupling/high temperature fixed point corresponding to have only the coefficient $\tilde \omega_0=1$ to be non--vanishing.

This difference can be expected already due to the different structure of the Migdal Kadanoff flow and the hierarchical flow. A typical feature of the Migdal Kadanoff flow is that the $\tilde \omega$ are convoluted with each other, which leads to a spreading effect of $\tilde \omega$ as a function of the representation labels. For the $S_3$ example it is therefore possible to flow from a configuration with $\tilde \omega_1=0$, i.e. where the sign representation does not appear, to the $BF$ fixed point, in which $\tilde \omega_1=1$. In the flow defined by $n-1$ Pachner moves this does not happen. The flow is described by vertex functions which depend on (six) representation labels. If initially the vertex functions are zero for any of the labels giving the sign (or any other) representation, this will be preserved by the flow. This is however a special feature of the $n-1$ Pachner moves. To get a more complete picture of the flow for more general lattice one has also to consider other Pachner moves, which will require some implementation of a truncation. The Migdal Kadanoff method is one particular truncation available for lattice gauge theories.  Other methods, such as tensor network renormalization \cite{levin, Dittrich:2011zh} are more general and in principle also applicable to spin foam models \cite{Dittrich:2011zh}. Tensor network renormalization can be motivated by using Pachner moves \cite{levin}, and indeed the Pachner moves different from the $n-1$ move are the ones where the truncation is implemented in tensor network renormalization.
This choice of a truncation can be also encoded into embedding maps which are closely related to the concept of cylindrical consistency in loop quantum gravity \cite{bd1205}.

\subsubsection{Renormalization group flow of an $S_3$ model via $1-4$ move}\label{p41}

In the following we will consider the flow of 3D models with group $S_3$ generated under the $1-4$ move.  We described this group in section \ref{s3ex}. We also choose $G' = \{1,a\}\simeq \mathbb{Z}_2$ as subgroup. 

We will define the model on the $2$-complex $\kappa_1$ dual to a tetrahedron. Its topology allows to glue four of them together to form the $2$-complex $\kappa_2$ dual to four tetrahedra which are the result of a 1-4 move. The boundary of that $2$-complex will be the same as the one dual to one tetrahedron, apart from trivial subdivisions of edges. The effective model on this, arising from integrating over internal degrees of freedom, can then be compared to the original model, by evaluating $Z_{\kappa_1}$ and $Z_{\kappa_2}$ on the same $S_3$-boundary spin network functions. Iterating this process generates a flow in the space of models.

If we define the model in terms of vertex functions, the flow can be carried out exactly. This is a special feature of all 1-$(n-1)$ moves, since they do not generate any non-localities. These moves have another interesting property, which is present for all models and dimensions: The vertex function can be interpreted as a function depending on spins and intertwiners. If one starts the flow with a function that is constrained to vanish whenever a specific spin or specific intertwiner is present, then this constraint is conserved under the 1-$(n-1)$ move. Therefore, if certain degrees of freedom are switched off, they stay switched off during the flow. This observation allows for a very convenient investigation of the RG flow, since it is automatically constrained to lie in certain submanifolds of the model space, all of which intersect in the high temperature fixed point. Note that this is a direct result of working on a hierarchic lattice, because the flow does not generate any non-localities. It is quite different from the hypercubic lattice flow described in the last section, where non-localities are truncated by the Migdal-Kadanoff approximation.\\[5pt]

Let us turn to $S_3$ once again. As previously described there are two $1$-dimensional representations (denoted $\mathbf{0}_{S_3}$ and $\mathbf{1}_{S_3}$), and one $2$-dimensional representation, denoted by $\mathbf{2}_{S_3}$. Since $\mathbf{0}_{S_3}$ is the trivial representation, the only nontrivial tensor products are given by
\begin{eqnarray*}
&&\mathbf{1}_{S_3}\otimes \mathbf{1}_{S_3}\;=\;\mathbf{0}_{S_3},\qquad\mathbf{1}_{S_3}\otimes \mathbf{2}_{S_3}\;=\;\mathbf{2}_{S_3}\\[5pt]
&&\mathbf{2}_{S_3}\otimes \mathbf{2}_{S_3}\;=\;\mathbf{0}_{S_3}\oplus \mathbf{1}_{S_3}\oplus \mathbf{2}_{S_3}
\end{eqnarray*}

\noindent The group $\mathbb{Z}_2$ has two $1$-dimensional representations, denoted by $\mathbf{0}_{\mathbb{Z}_2}$ and $\mathbf{1}_{\mathbb{Z}_2}$, satisfying $\mathbf{1}_{\mathbb{Z}_2}\otimes \mathbf{1}_{\mathbb{Z}_2}=\mathbf{0}_{\mathbb{Z}_2}$. Decomposing the representation spaces of $S_3$ with respect to $\mathbb{Z}_2$ yields
\begin{eqnarray}\nonumber
\mathbf{0}_{S_3}\;&=&\;\mathbf{0}_{\mathbb{Z}_2}\\[5pt]\label{Eq:DecompositionOfS_3IntoZ_2}
\mathbf{1}_{S_3}\;&=&\;\mathbf{1}_{\mathbb{Z}_2}\\[5pt]\nonumber
\mathbf{2}_{S_3}\;&=&\;\mathbf{0}_{\mathbb{Z}_2}\oplus \mathbf{1}_{\mathbb{Z}_2}
\end{eqnarray}

\noindent After gauge-invariance and discrete symmetries of the tetrahedron, there are eleven gauge-invariant combinations of $S_3$-irreps along the edges of a tetrahedron. In other words, a vertex function is given by specifying the eleven different values of the $\{6j\}$-symbol for $S_3$.\footnote{Face- and edge amplitudes can in principle be absorbed into the vertex functions in this case, so we will fix them to equal those of the $S_3$ BF theory, for convenience.}

If we specify a model in terms of $E$-functions, then there are fewer parameters: Due to the decomposition (\ref{Eq:DecompositionOfS_3IntoZ_2}) the most general $E$-function is -- after normalization -- of the form
\begin{eqnarray}\label{Eq:S_3ModelInTermsOfEFunctions}
E(h)\;=\;1\,+\,e_1\,\text{sgn}(h)\,+2\,e_{20}\,\rho_{\mathbf{2}_{S_3}}(h)_{00}\,+\,\, 2\,e_{21}\,\rho_{\mathbf{2}_{S_3}}(h)_{11}
\end{eqnarray}

\noindent where $\text{sgn}(h)=\text{tr}\rho_{\mathbf{1}_{S_3}}(h)$ is the signum of the permutation $h\in S_3$, and $(00)$ and $(11)$ specify the upper left and the lower right entry of the $2\times 2$-matrix $\rho_{\mathbf{2}_{S_3}}(h)$. A model is then given by the three parameters $e_1$, $e_{20}$ and $e_{21}$.\footnote{ These parameters are connected to the parameters in (\ref{gpar}) by $\beta_1=1+e_1+2(e_{20}+e_{21}),\,\beta_2=1-e_1+2(e_{20}-e_{21}),\, \beta_3=1-e_1+4(e_{20}-e_{21}),\,\beta_4=1+e_1-(e_{20}-e_{21})$.}

Note that every model in terms of $E$-functions can be rewritten in terms of vertex functions, but since these are more general, the converse is not necessarily true. Moreover, it turns out that the RG flow does not preserve the class of models of the form (\ref{Eq:S_3ModelInTermsOfEFunctions}), but takes place in the eleven-dimensional space of all vertex functions instead. This puts us into the favourable position of comparing the the exact flow to the truncated flow, which is given by projecting onto the space of models of the form (\ref{Eq:S_3ModelInTermsOfEFunctions}) after each step.\footnote{There is no unique way of projecting, but different choices do not seem to yield qualitatively different results.} In the following we present the results.\\

First of all, there are three obvious fixed points, which exist both under the exact and the truncated flow. All three are models which can be expressed in terms of $E$-functions as in (\ref{Eq:S_3ModelInTermsOfEFunctions}). In these parameters, they are given by
\begin{itemize}
\item [I)] $S_3$ BF theory: $e_1=e_{20}=e_{21}=1$
\item [II)] $\mathbf{Z}_2$-BF theory: $e_1 = 1$, $e_{20} = e_{21} = 0$
\item [III)] High-temperature fixed point: $e_1=e_{20}=e_{21}=0$
\end{itemize}

\noindent Point $\text{I)}$ has only unstable directions -- a slight derivation will either lead to the model flowing to diverging couplings or into one of the other two fixed points.
In point $\text{II)}$ all couplings associated to terms containing $\mathbf{2}_{S_3}$ are turned off. Because of (\ref{Eq:DecompositionOfS_3IntoZ_2}), the two remaining irreps of $S_3$ can be interpreted as irreps of $\mathbf{Z}_2$, so the point corresponds to $\mathbb{Z}_2$-BF theory. This point has both stable and unstable directions. In point $\text{III)}$, which is completely attractive, also $\mathbf{1}_{S_3}$ is switched off. The resulting theory can be either interpreted as BF theory for the trivial group, or -- in the language of measure theory -- as the Ashtekar-Lewandowski measure for $S_{3}$ functions on the boundary. On it the kinematical and physical Hilbert space coincide, and in statistical field theory, it corresponds to the high temperature fixed point (HTF).\\

Looking at the exact flow equations for the eleven parameters of the vertex functions, one can find, however, several more fixed points. These do not correspond to any topological theory, but can be interpreted as nontrivial. One of the points, which is rather close to the other three fixed points, is given, in terms of the vertex function couplings (i.e.~the values of the $\{6j\}$-symbols) all being zero, apart from
\begin{eqnarray*}
\alpha_1\;=\;\SixJ{\mathbf{0}_{S_3}}{\mathbf{0}_{S_3}}{\mathbf{0}_{S_3}}{\mathbf{1}_{S_3}}{\mathbf{1}_{S_3}}{\mathbf{1}_{S_3}},\qquad
\alpha_2\;=\;\SixJ{\mathbf{0}_{S_3}}{\mathbf{1}_{S_3}}{\mathbf{1}_{S_3}}{\mathbf{0}_{S_3}}{\mathbf{1}_{S_3}}{\mathbf{1}_{S_3}}
\end{eqnarray*}

\begin{figure}[hbt]
\begin{center}
	\psfrag{e1}{$e_1$}
    \psfrag{e20}{$e_{20}$}
    \psfrag{e21}{$e_{21}$}
    \psfrag{p1}{I}
    \psfrag{p2}{II}
    \psfrag{p3}{III}
    \psfrag{p4}{IV$^*$}
\includegraphics[scale=0.40]{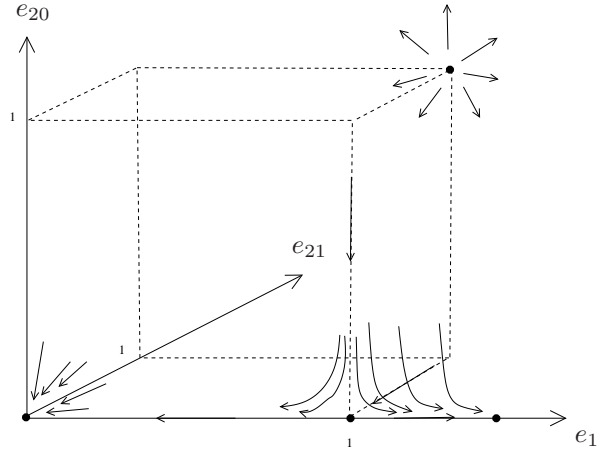}
\caption{Truncated RG flow of $S_3$ model under $1-4$ move. The diagram includes four fixed points of the flow. }\label{Fig:RGFlow}
\end{center}
\end{figure}

\noindent The exact recursion equations for this system are given by
\begin{eqnarray}\label{Eq:RecursionEquationsAlcoholicFixedPoint}
\alpha_1\;\longrightarrow\;\frac{\alpha_1^3+\alpha_1\alpha_2^3}{1+\alpha_1^4},\qquad
\alpha_2\;\longrightarrow\;\frac{2\alpha_1^2\alpha_2^2}{1+\alpha_1^4}
\end{eqnarray}

\noindent The obvious fixed point solution is given by $\alpha_1=\alpha_2=1$, which corresponds to point $\text{II)}$. Another nontrivial solution is e.g.~given by
\begin{itemize}
\item[IV)] $\alpha_1=2.54605,\;\alpha_2=3.31832$
\end{itemize}
\noindent The values for $\alpha_1$ and $\alpha_2$ can be expressed as roots of polynomials of order five, and we are not aware of a closed expression. The point $\text{IV)}$ seems to have several attractive directions, and the correct perturbation from point $\text{I)}$ can let the flow run into point $\text{IV)}$. The point is \emph{not} topological, in particular it defines a model which is not invariant under trivial subdivisions of bulk edges by a $2$-valent vertex, unlike all the BF theories. Also, the resulting model is \emph{not} invariant under the $2-3$ move, as one can readily check.

Since the models given by (\ref{Eq:S_3ModelInTermsOfEFunctions}) always lead to coefficients of the vertex functions satisfying $\alpha_1=e_1^3$ and $\alpha_2=e_1^4$, one can readily check that the point $\text{IV)}$ is not of the form (\ref{Eq:S_3ModelInTermsOfEFunctions}). It lies, however, quite close to the three-dimensional submanifold of models given by an $E$-function. \\

Under the truncated flow, in which one projects onto the space of $E$-function models as in (\ref{Eq:S_3ModelInTermsOfEFunctions}) after each step, one obtains a flow in the three-dimensional space of parameters $(e_1,e_{20},e_{21})$. One does not only recover the three obvious fixed points $\text{I)}$, $\text{II)}$ and $\text{III)}$, but remarkably there is another fixed point which lies at roughly

\begin{itemize}
\item[IV${}^*$)] $e_1=1.4021$, $e_{20}=e_{21}= 0$
\end{itemize}

\noindent It can be readily checked that the vertex function derived from fixed point $\text{IV}^*\text{)}$, which has the only nonzero couplings $\alpha_1=e_1^3$ and $\alpha_2=e_1^4$, is rather close to the fixed point under the exact flow $\text{IV})$. This demonstrates that truncating the theory to models of the form (\ref{Eq:S_3ModelInTermsOfEFunctions}) can reveal characteristic nontrivial features of the phase space.

\subsection{Outlook}\label{out}

In particular the example in the last section gives as hope that the parametrization with $E$ function presented in this work captures the relevant parameter for the renormalization flow. More generally we have seen that there are a number of features similar to the behaviour of lattice gauge theories \cite{Migdal:1975zg}, in particular the appearance of the  strong coupling/ high temperature fixed point and the $BF$ (low temperature) fixed points. (If there is a normal subgroup, a $BF$ theory on the quotient group will also define a fixed point.) The $BF$ fixed points are unstable in 2D and 3D, whereas the high temperature fixed points has attractive directions. It is conjectured that non-Abelian lattice gauge theories do not show phase transitions in 4D, hence that as in 3D all models flow to the high temperature fixed point.

The spin foam models provide a larger phase space and a crucial question for future research will be if this larger phase space allows for additional phases/ fixed points. This question can be investigated with spin nets models \cite{Bahr:2011yc,Dittrich:2011zh}, which are `dimensionally reduced spin foams'.  Interestingly spin nets can be formulated with the same algebraic ingredients as spin foams, i.e. any spin foam model formulated with $E$ functions leads to a corresponding spin net model. The advantage is that these models are already non--trivial in two dimensions which simplifies enormously coarse graining considerations. Additionally, it is known that 4D lattice gauge theories show a similar flow behaviour as the corresponding 2D (spin net/edge ) models \cite{Kadanoff:1976jb}. If this similarity extends to  spin foam models, these investigations in 2D could already give information on the 4D spin foam models.

In this work we discussed  $n-1$ Pachner moves,  whose corresponding renormalization flow can be obtained exactly if vertex amplitudes are used. Classical  Regge calculus is invariant under $n-1$ Pachner moves, but 4D Regge calculus is not invariant under the $3-3$ move \cite{sebastian}. Finding theories (fixed points) invariant under the $n-1$ Pachner moves could therefore lead to a quantum theory mimicking better the symmetry properties of the classical one.

On the other hand  the flow generated by the $n-1$ Pachner moves is quite special. It leads not only to a closed, exact flow for the vertex amplitudes but also features invariant submanifolds, resulting from the set of vertex amplitudes that are zero if a given  representation appears as argument. Correspondingly one can find a number of fixed points for this flow, which do not appear i.e. in the Migdal Kadanoff scheme for lattice gauge theories.  The question arises, whether these fixed points have any significance, i.e. would flow to other fixed points, if we consider more general coarse grainings.

 Therefore also the other Pachner moves have to be considered, or alternatively the blocking of a regular lattice, which incorporates the other Pachner moves. To this end truncations have to be used, as otherwise the number of coupling parameters will grow without bound. A truncation which is suggested by this work, is given by vertex amplitudes that can be formulated through $E$--functions. Again the main question will be whether this truncation captures the relevant (local) couplings.

In this work we considered either coarse graining procedures which only lead to local couplings (two dimensional theories, $n-1$ Pachner moves) or a truncation of the flow to local couplings (Migdal Kadanoff procedure). For some questions, like the restoration of diffeomorphism symmetry via coarse graining \cite{he,DittrichZak}, or in order  to obtain a more reliable phase diagram, the consideration of non-local couplings is necessary. 
 The difficulty in dealing with non-local couplings is that these will proliferate and are hard to interpret and to control. 

One way to give an interpretation of  truncations to parameter spaces including non-local couplings, is to introduce non--elementary building blocks which carry more boundary data than the building blocks one starts with \cite{bd1205}. The growing number of coupling parameters is now reflected in the growing number of boundary data, between which the couplings can occur. These couplings can be encoded into a functional which acts on the boundary Hilbert space of the (non--elementary) building block in question. The universal boundary Hilbert space defined in section \ref{hilb} provides in this sense a more general parameter space, describing also models with non--local couplings. (Here non--local is to be understood to include more than just the nearest neighbour couplings.) 
Again to obtain a closed coarse graining flow and not a flow which rather maps to larger and larger Hilbert spaces (leading to the continuum limit, i.e. a path integral over a region with continuum boundary data), one has to implement truncations. Such truncations can be described by maps \cite{bd1205} which embed the boundary Hilbert spaces into each other.

\section{Summary}\label{sum}

In this work we proposed a new holonomy formulation for spin foam models, which is able to describe all\footnote{With the exception of the KKL model \cite{Kaminski:2009fm}.} the current models. This class of models features, compared to other formulations, a minimal set of parameters, and is naturally extendable to finite groups and arbitrary two--complexes. %Boundary Hilbert space can be naturally described and are one method to clarify the relations with other formulations.
We related this formulation to other holonomy formulations and the operator %as well as spin (network)
formulation of spin foams. Further applications of this holonomy formulation will appear in \cite{HS,SL}

%Also the holonomy formulation allows to generalize spin foam models to other groups, in particular finite ones.

The choice between different formulation\footnote{
Holonomy and spin representations underly in lattice gauge theory the weak and strong coupling expansion respectively. Hence these different representations help to identify the appropriate degrees of freedom in the different regimes. For pure gravitational spin foam models such a distinction of regimes is much less clear, but could corresponds to a phase where geometry is mostly degenerate (strong coupling) and a (near) topological phase  (weak coupling).}
of a class of models becomes most relevant for coarse graining and the construction of renormalization group flows, as it defines the space of models in which the renormalization flow takes place (or to which it is projected/truncated). Here the question is still open which spin foam formulation is the most appropriate, at least for the four--dimensional case.

The holonomy formulation suggest to implement coarse graining techniques from lattice gauge theories, and we followed this strategy for the 2D and 3D dimensional theories, that are in a certain sense equivalent to gauge theories. On the other hand spin foams have a richer structure than standard lattice gauge theories: in the $E$ function formulation introduced here this can be made precise, as in spin foams  $E$ functions  are allowed which are not  class functions.
The main question for future research will be how relevant under coarse graining the parameters are that describe the deviations of the $E$ functions from a class function. This will reveal whether the algebraic structures specific to spin foams could lead to macroscopic phases different from those of standard lattice gauge theory.

%Coarse graining techniques with a clear geometric interpretation, such as Pachner moves might be more appropriate.

\appendix

\section{Non-factorizable model for a 3D dual simplicial lattice}
\label{app-example}

We will prove in this appendix that not all
3-valent edge $C$ functions coming from $E$ have the factorization property (\ref{eq-factor}).

For a three valent edge in case of $SU(2)\subset Spin(4)$ we know that every $C$
function can be
written in the form
\begin{equation}
 C(g_1,g_2,g_3)=\sum_{j^\pm_i} \tilde{C}_{j^+_1,j^-_1,j^+_2,j^-_2,j^+_3,j^-_3}\
\psi_{j^\pm_i}(g_1,g_2,g_3)
\end{equation}
where $g_i\in Spin(4)$ and
\begin{equation}
 \psi_{j^\pm_i}=\iota_{j^\pm_i}^{m_1m_2m_3}D_{j_1}(g_1)^{m_1'}_{m_1}
D_{j_2}(g_2)^{m_2'}_{m_2}D_{j_3}(g_3)^{m_3'}_{m_3}\iota_{j^\pm_i\ m_1'm_2'm_3'}
\end{equation}
and Clebsches $\iota$ are normalized.

Suppose that we are starting from the $E$ function with some $e_{(j^+j^-)k}$ be
equal
\begin{equation}
 \begin{split}
&e_{(j,j)0}=1\ \text{for}\ j\not=1\\
 &e_{(1,1)1}=1
 \end{split}
\end{equation}
and otherwise $0$.

Then we can compute
$\tilde{C}_{j^\pm_i}$ from $E$ functions using \eqref{d06gen4} as follows
\begin{equation}
 \tilde{C}_{j^\pm_i}=\left(\prod_i e_{(j^+_ij^-_i)k_i}\right)
\langle\iota_{j^\pm_i}|\prod_i
P_{(j^+_i,j^-_i)k_i}|\iota_{j^\pm_i}\rangle
\end{equation}
where $k_i$ are the only $k$ when $e$ is nonzero and $P_{j^+j^-k}$ is the
projection onto $SU(2)$ subrepresentation with spin $k$.
This in fact is the product of
$\prod_i
e_{(j^+_ij^-_i)k_i}$ and a square of suitable contraction of Clebsches since
\begin{equation}
\langle\iota_{j^\pm_i}|\prod_i
P_{(j^+_i,j^-_i)k_i}|\iota_{j^\pm_i}\rangle=
 \left|\iota_{j^\pm_i}\left(\prod_i
\iota_{j^+_i,j^-_ik_i}\right)\iota_{k_1k_2k_3}\right|^2
\end{equation}

One can check the following properties:
\begin{itemize}
 \item[1] if $j_i\not=1$ and $(j_1,j_2,j_3)$
satisfied triangle inequalities, $\sum j_i$ is an integer then
\begin{equation}
\tilde{C}_{j_1,j_1,j_2,j_2,j_3,j_3}\not=0\ . \label{eq1}
\end{equation}
\item[2] we have
\begin{equation}
 \tilde{C}_{j_1,j_1,j_2,j_2,1,1}=0\ ,\label{eq2}
\end{equation}
for $j_i\not=1$ because $k_i$ do not satisfy triangle
inequality.
\item[3] We have
\begin{equation}
 \tilde{C}_{2,2,1,1,1,1}\not=0\ . \label{eq3}
\end{equation}
proved by use of diagrams.
\end{itemize}

Suppose that $\tilde{C}_{j_i^\pm}$ has the factorization property, then it can
be written in the form
\begin{equation}
 \tilde{C}_{j_1,j_1,j_2,j_2,j_3,j_3}=\epsilon_{j_1,j_1}\epsilon_{j_2,j_2}
\epsilon_{j_3, j_3}
\end{equation}
but from \eqref{eq1} we know that
\begin{equation}
 \epsilon_{j,j}\not=0\ \text{for}\ j\not=1\ .
\end{equation}
By the property \eqref{eq2} we thus obtain
\begin{equation}
 \epsilon_{1,1}=0
\end{equation}
but then \eqref{eq3} cannot be satisfied.

We thus proved that for this $E$ the function $C$ cannot be written in the
factorized form.

\vspace{1cm}
\section*{Acknowledgements}

BD thanks Valentin Bonzom for discussions.
Research at Perimeter Institute is supported by the Government of Canada through Industry Canada and by the Province of Ontario through the Ministry of Research and Innovation.

\footnotesize{

\bibliographystyle{utphys} %\bibliographystyle{iopart-num}   %{unsrt}   %the first gives no titles, the second no arxiv ref's
\bibliography{HolSFM}

\providecommand{\href}[2]{#2}\begingroup\raggedright\begin{thebibliography}{10}

\bibitem{bookcarlo}
C.~Rovelli, ``{Zakopane lectures on loop gravity},''
  \href{http://arxiv.org/abs/1102.3660}{{\ttfamily arXiv:1102.3660 [gr-qc]}}.

\bibitem{alexreview}
A.~Perez, ``{The Spin Foam Approach to Quantum Gravity},''
\href{http://arxiv.org/abs/1205.2019}{{\ttfamily arXiv:1205.2019 [gr-qc]}}.
%%CITATION = ARXIV:1205.2019;%%.

\bibitem{smolin}
C.~Rovelli and L.~Smolin, ``{Discreteness of area and volume in quantum
  gravity},'' \href{http://dx.doi.org/10.1016/0550-3213(95)00150-Q,
  10.1016/0550-3213(95)00150-Q}{{\em Nucl.Phys.} {\bfseries B442} (1995)
  593--622}, \href{http://arxiv.org/abs/gr-qc/9411005}{{\ttfamily
  arXiv:gr-qc/9411005 [gr-qc]}}.

\bibitem{ash}
A.~Ashtekar and J.~Lewandowski, ``{Quantum theory of geometry. 1: Area
  operators},'' \href{http://dx.doi.org/10.1088/0264-9381/14/1A/006}{{\em
  Class.Quant.Grav.} {\bfseries 14} (1997) A55--A82},
  \href{http://arxiv.org/abs/gr-qc/9602046}{{\ttfamily arXiv:gr-qc/9602046
  [gr-qc]}}.

\bibitem{Barrett:1998gs}
J.~W. Barrett and R.~M. Williams, ``{The Asymptotics of an amplitude for the
  four simplex},'' {\em Adv.Theor.Math.Phys.} {\bfseries 3} (1999) 209--215,
\href{http://arxiv.org/abs/gr-qc/9809032}{{\ttfamily arXiv:gr-qc/9809032
  [gr-qc]}}.
%%CITATION = GR-QC/9809032;%%.

\bibitem{Barrett:2010ex}
J.~W. Barrett, R.~Dowdall, W.~J. Fairbairn, H.~Gomes, F.~Hellmann, {\em
  et~al.}, ``{Asymptotics of 4d spin foam models},''
  \href{http://dx.doi.org/10.1007/s10714-010-0983-7}{{\em Gen.Rel.Grav.}
  {\bfseries 43} (2011) 2421--2436},
\href{http://arxiv.org/abs/1003.1886}{{\ttfamily arXiv:1003.1886 [gr-qc]}}.
%%CITATION = ARXIV:1003.1886;%%.

\bibitem{Conrady:2008mk}
F.~Conrady and L.~Freidel, ``{On the semiclassical limit of 4d spin foam
  models},'' \href{http://dx.doi.org/10.1103/PhysRevD.78.104023}{{\em
  Phys.Rev.} {\bfseries D78} (2008) 104023},
\href{http://arxiv.org/abs/0809.2280}{{\ttfamily arXiv:0809.2280 [gr-qc]}}.
%%CITATION = ARXIV:0809.2280;%%.

\bibitem{Dittrich:2008pw}
B.~Dittrich, ``{Diffeomorphism symmetry in quantum gravity models},''
\href{http://arxiv.org/abs/0810.3594}{{\ttfamily arXiv:0810.3594 [gr-qc]}}.
%%CITATION = ARXIV:0810.3594;%%.

\bibitem{Bahr:2009ku}
B.~Bahr and B.~Dittrich, ``{(Broken) Gauge Symmetries and Constraints in Regge
  Calculus},'' \href{http://dx.doi.org/10.1088/0264-9381/26/22/225011}{{\em
  Class.Quant.Grav.} {\bfseries 26} (2009) 225011},
\href{http://arxiv.org/abs/0905.1670}{{\ttfamily arXiv:0905.1670 [gr-qc]}}.
%%CITATION = ARXIV:0905.1670;%%.

\bibitem{Bahr:2009qc}
B.~Bahr and B.~Dittrich, ``{Improved and Perfect Actions in Discrete
  Gravity},'' \href{http://dx.doi.org/10.1103/PhysRevD.80.124030}{{\em
  Phys.Rev.} {\bfseries D80} (2009) 124030},
\href{http://arxiv.org/abs/0907.4323}{{\ttfamily arXiv:0907.4323 [gr-qc]}}.
%%CITATION = ARXIV:0907.4323;%%.

\bibitem{Rovelli:2011fk}
C.~Rovelli, ``{Discretizing parametrized systems: the magic of
  Ditt-invariance},''
\href{http://arxiv.org/abs/1107.2310}{{\ttfamily arXiv:1107.2310 [hep-lat]}}.
%%CITATION = ARXIV:1107.2310;%%.

\bibitem{DittrichZak}
B.~Dittrich, ``{How to construct diffeomorphism symmetry on the lattice},''
\href{http://arxiv.org/abs/1201.3840}{{\ttfamily arXiv:1201.3840 [gr-qc]}}.
%%CITATION = ARXIV:1201.3840;%%.

\bibitem{fotini}
F.~Markopoulou, ``{Coarse graining in spin foam models},''
  \href{http://dx.doi.org/10.1088/0264-9381/20/5/301}{{\em Class.Quant.Grav.}
  {\bfseries 20} (2003) 777--800},
  \href{http://arxiv.org/abs/gr-qc/0203036}{{\ttfamily arXiv:gr-qc/0203036
  [gr-qc]}}.

\bibitem{oeckl}
R.~Oeckl, ``{Renormalization of discrete models without background},''
  \href{http://dx.doi.org/10.1016/S0550-3213(03)00145-7}{{\em Nucl.Phys.}
  {\bfseries B657} (2003) 107--138},
  \href{http://arxiv.org/abs/gr-qc/0212047}{{\ttfamily arXiv:gr-qc/0212047
  [gr-qc]}}.

\bibitem{Dittrich:2011zh}
B.~Dittrich, F.~C. Eckert, and M.~Martin-Benito, ``{Coarse graining methods for
  spin net and spin foam models},''
  \href{http://dx.doi.org/10.1088/1367-2630/14/3/035008}{{\em New J.Phys.}
  {\bfseries 14} (2012) 035008},
\href{http://arxiv.org/abs/1109.4927}{{\ttfamily arXiv:1109.4927 [gr-qc]}}.
%%CITATION = ARXIV:1109.4927;%%.

\bibitem{Dittrich:2011av}
B.~Dittrich and F.~C. Eckert, ``{Towards computational insights into the
  large-scale structure of spin foams},''
  \href{http://dx.doi.org/10.1088/1742-6596/360/1/012004}{{\em
  J.Phys.Conf.Ser.} {\bfseries 360} (2012) 012004},
\href{http://arxiv.org/abs/1111.0967}{{\ttfamily arXiv:1111.0967 [gr-qc]}}.
%%CITATION = ARXIV:1111.0967;%%.

\bibitem{Barrett:1997gw}
J.~W. Barrett and L.~Crane, ``{Relativistic spin networks and quantum
  gravity},'' \href{http://dx.doi.org/10.1063/1.532254}{{\em J.Math.Phys.}
  {\bfseries 39} (1998) 3296--3302},
\href{http://arxiv.org/abs/gr-qc/9709028}{{\ttfamily arXiv:gr-qc/9709028
  [gr-qc]}}.
%%CITATION = GR-QC/9709028;%%.

\bibitem{Engle:2007wy}
J.~Engle, E.~Livine, R.~Pereira, and C.~Rovelli, ``{LQG vertex with finite
  Immirzi parameter},''
  \href{http://dx.doi.org/10.1016/j.nuclphysb.2008.02.018}{{\em Nucl.Phys.}
  {\bfseries B799} (2008) 136--149},
\href{http://arxiv.org/abs/0711.0146}{{\ttfamily arXiv:0711.0146 [gr-qc]}}.
%%CITATION = ARXIV:0711.0146;%%.

\bibitem{Freidel:2007py}
L.~Freidel and K.~Krasnov, ``{A New Spin Foam Model for 4d Gravity},''
  \href{http://dx.doi.org/10.1088/0264-9381/25/12/125018}{{\em
  Class.Quant.Grav.} {\bfseries 25} (2008) 125018},
\href{http://arxiv.org/abs/0708.1595}{{\ttfamily arXiv:0708.1595 [gr-qc]}}.
%%CITATION = ARXIV:0708.1595;%%.

\bibitem{Bahr:2011yc}
B.~Bahr, B.~Dittrich, and J.~P. Ryan, ``{Spin foam models with finite
  groups},''
\href{http://arxiv.org/abs/1103.6264}{{\ttfamily arXiv:1103.6264 [gr-qc]}}.
%%CITATION = ARXIV:1103.6264;%%.

\bibitem{PFEIF}
H.~Pfeiffer, ``{Dual variables and a connection picture for the Euclidean
  Barrett-Crane model},''
  \href{http://dx.doi.org/10.1088/0264-9381/19/6/306}{{\em Class.Quant.Grav.}
  {\bfseries 19} (2002) 1109--1138},
  \href{http://arxiv.org/abs/gr-qc/0112002}{{\ttfamily arXiv:gr-qc/0112002
  [gr-qc]}}.

\bibitem{val}
V.~Bonzom and E.~R. Livine, ``{A Lagrangian approach to the Barrett-Crane spin
  foam model},'' \href{http://dx.doi.org/10.1103/PhysRevD.79.064034}{{\em
  Phys.Rev.} {\bfseries D79} (2009) 064034},
\href{http://arxiv.org/abs/0812.3456}{{\ttfamily arXiv:0812.3456 [gr-qc]}}.
%%CITATION = ARXIV:0812.3456;%%.

\bibitem{Magliaro:2010ih}
E.~Magliaro and C.~Perini, ``{Local spin foams},''
\href{http://arxiv.org/abs/1010.5227}{{\ttfamily arXiv:1010.5227 [gr-qc]}}.
%%CITATION = ARXIV:1010.5227;%%.

\bibitem{benloops}
B.~Bahr, ``{Operator Spin Foams: holonomy formulation and coarse graining},''
  \href{http://dx.doi.org/10.1088/1742-6596/360/1/012042}{{\em
  J.Phys.Conf.Ser.} {\bfseries 360} (2012) 012042},
\href{http://arxiv.org/abs/1112.3567}{{\ttfamily arXiv:1112.3567 [gr-qc]}}.
%%CITATION = ARXIV:1112.3567;%%.

\bibitem{Bahr:2010bs}
B.~Bahr, F.~Hellmann, W.~Kaminski, M.~Kisielowski, and J.~Lewandowski,
  ``{Operator Spin Foam Models},''
  \href{http://dx.doi.org/10.1088/0264-9381/28/10/105003}{{\em
  Class.Quant.Grav.} {\bfseries 28} (2011) 105003},
\href{http://arxiv.org/abs/1010.4787}{{\ttfamily arXiv:1010.4787 [gr-qc]}}.
%%CITATION = ARXIV:1010.4787;%%.

\bibitem{Alexandrov:2002xc}
S.~Alexandrov, ``{Hilbert space structure of covariant loop quantum gravity},''
  \href{http://dx.doi.org/10.1103/PhysRevD.66.024028}{{\em Phys.Rev.}
  {\bfseries D66} (2002) 024028},
\href{http://arxiv.org/abs/gr-qc/0201087}{{\ttfamily arXiv:gr-qc/0201087
  [gr-qc]}}.
%%CITATION = GR-QC/0201087;%%.

\bibitem{Livine:2002ak}
E.~R. Livine, ``{Projected spin networks for Lorentz connection: Linking spin
  foams and loop gravity},''
  \href{http://dx.doi.org/10.1088/0264-9381/19/21/316}{{\em Class.Quant.Grav.}
  {\bfseries 19} (2002) 5525--5542},
\href{http://arxiv.org/abs/gr-qc/0207084}{{\ttfamily arXiv:gr-qc/0207084
  [gr-qc]}}.
%%CITATION = GR-QC/0207084;%%.

\bibitem{HS}
B.~Bahr, B.~Dittrich, F.~Hellmann, and W.~Kaminski, ``{Holonomy Spin Foam
  Models: Boundary Hilbert spaces and canonical dynamics},''  (2012) .

\bibitem{SL}
F.~Hellmann and W.~Kaminski, ``{Holonomy Spin Foam Models: Asymptotic dynamics of EPRL type models},''
  (2012) .

\bibitem{val2}
V.~Bonzom, ``{From lattice BF gauge theory to area-angle Regge calculus},''
  \href{http://dx.doi.org/10.1088/0264-9381/26/15/155020}{{\em
  Class.Quant.Grav.} {\bfseries 26} (2009) 155020},
\href{http://arxiv.org/abs/0903.0267}{{\ttfamily arXiv:0903.0267 [gr-qc]}}.
%%CITATION = ARXIV:0903.0267;%%.

\bibitem{aa}
B.~Dittrich and S.~Speziale, ``{Area-angle variables for general relativity},''
  \href{http://dx.doi.org/10.1088/1367-2630/10/8/083006}{{\em New J.Phys.}
  {\bfseries 10} (2008) 083006},
\href{http://arxiv.org/abs/0802.0864}{{\ttfamily arXiv:0802.0864 [gr-qc]}}.
%%CITATION = ARXIV:0802.0864;%%.

\bibitem{Bonzom:2010ar}
V.~Bonzom and M.~Smerlak, ``{Bubble divergences from cellular cohomology},''
  \href{http://dx.doi.org/10.1007/s11005-010-0414-4}{{\em Lett.Math.Phys.}
  {\bfseries 93} (2010) 295--305},
\href{http://arxiv.org/abs/1004.5196}{{\ttfamily arXiv:1004.5196 [gr-qc]}}.
%%CITATION = ARXIV:1004.5196;%%.

\bibitem{DePietri:1999bx}
R.~De~Pietri, L.~Freidel, K.~Krasnov, and C.~Rovelli, ``{Barrett-Crane model
  from a Boulatov-Ooguri field theory over a homogeneous space},''
  \href{http://dx.doi.org/10.1016/S0550-3213(00)00005-5}{{\em Nucl.Phys.}
  {\bfseries B574} (2000) 785--806},
\href{http://arxiv.org/abs/hep-th/9907154}{{\ttfamily arXiv:hep-th/9907154
  [hep-th]}}.
%%CITATION = HEP-TH/9907154;%%.

\bibitem{Reisenberger:2000zc}
M.~P. Reisenberger and C.~Rovelli, ``{Space-time as a Feynman diagram: The
  Connection formulation},''
  \href{http://dx.doi.org/10.1088/0264-9381/18/1/308}{{\em Class.Quant.Grav.}
  {\bfseries 18} (2001) 121--140},
\href{http://arxiv.org/abs/gr-qc/0002095}{{\ttfamily arXiv:gr-qc/0002095
  [gr-qc]}}.
%%CITATION = GR-QC/0002095;%%.

\bibitem{Geloun:2010vj}
J.~Ben~Geloun, R.~Gurau, and V.~Rivasseau, ``{EPRL/FK Group Field Theory},''
  \href{http://dx.doi.org/10.1209/0295-5075/92/60008}{{\em Europhys.Lett.}
  {\bfseries 92} (2010) 60008},
\href{http://arxiv.org/abs/1008.0354}{{\ttfamily arXiv:1008.0354 [hep-th]}}.
%%CITATION = ARXIV:1008.0354;%%.

\bibitem{oritigft}
D.~Oriti, ``{The Group field theory approach to quantum gravity},''
\href{http://arxiv.org/abs/gr-qc/0607032}{{\ttfamily arXiv:gr-qc/0607032
  [gr-qc]}}.
%%CITATION = GR-QC/0607032;%%.

\bibitem{Gurau:2009tw}
R.~Gurau, ``{Colored Group Field Theory},''
  \href{http://dx.doi.org/10.1007/s00220-011-1226-9}{{\em Commun.Math.Phys.}
  {\bfseries 304} (2011) 69--93},
\href{http://arxiv.org/abs/0907.2582}{{\ttfamily arXiv:0907.2582 [hep-th]}}.
%%CITATION = ARXIV:0907.2582;%%.

\bibitem{etsim}
E.~R. Livine and S.~Speziale, ``{A New spinfoam vertex for quantum gravity},''
  \href{http://dx.doi.org/10.1103/PhysRevD.76.084028}{{\em Phys.Rev.}
  {\bfseries D76} (2007) 084028},
\href{http://arxiv.org/abs/0705.0674}{{\ttfamily arXiv:0705.0674 [gr-qc]}}.
%%CITATION = ARXIV:0705.0674;%%.

\bibitem{Migdal:1975zg}
A.~A. Migdal, ``{Recursion Equations in Gauge Theories},''
{\em Sov.Phys.JETP} {\bfseries 42} (1975) 413.
%%CITATION = SPHJA,42,413;%%.

\bibitem{Ben}
B.~Bahr, ``{On knottings in the physical Hilbert space of LQG as given by the
  EPRL model},'' \href{http://dx.doi.org/10.1088/0264-9381/28/4/045002}{{\em
  Class.Quant.Grav.} {\bfseries 28} (2011) 045002},
\href{http://arxiv.org/abs/1006.0700}{{\ttfamily arXiv:1006.0700 [gr-qc]}}.
%%CITATION = ARXIV:1006.0700;%%.

\bibitem{Bojowald:2009im}
M.~Bojowald and A.~Perez, ``{Spin foam quantization and anomalies},''
  \href{http://dx.doi.org/10.1007/s10714-009-0892-9}{{\em Gen.Rel.Grav.}
  {\bfseries 42} (2010) 877--907},
\href{http://arxiv.org/abs/gr-qc/0303026}{{\ttfamily arXiv:gr-qc/0303026
  [gr-qc]}}.
%%CITATION = GR-QC/0303026;%%.

\bibitem{Bahr:2011uj}
B.~Bahr, B.~Dittrich, and S.~Steinhaus, ``{Perfect discretization of
  reparametrization invariant path integrals},''
  \href{http://dx.doi.org/10.1103/PhysRevD.83.105026}{{\em Phys.Rev.}
  {\bfseries D83} (2011) 105026},
\href{http://arxiv.org/abs/1101.4775}{{\ttfamily arXiv:1101.4775 [gr-qc]}}.
%%CITATION = ARXIV:1101.4775;%%.

\bibitem{Kaminski:2009fm}
W.~Kaminski, M.~Kisielowski, and J.~Lewandowski, ``{Spin-Foams for All Loop
  Quantum Gravity},'' \href{http://dx.doi.org/10.1088/0264-9381/29/4/049502,
  10.1088/0264-9381/27/9/095006}{{\em Class.Quant.Grav.} {\bfseries 27} (2010)
  095006},
\href{http://arxiv.org/abs/0909.0939}{{\ttfamily arXiv:0909.0939 [gr-qc]}}.
%%CITATION = ARXIV:0909.0939;%%.

\bibitem{Kadanoff:1976jb}
L.~Kadanoff, ``{Notes on Migdal's Recursion Formulas},''
\href{http://dx.doi.org/10.1016/0003-4916(76)90066-X}{{\em Annals Phys.}
  {\bfseries 100} (1976) 359--394}.
%%CITATION = APNYA,100,359;%%.

\bibitem{Ito}
K.~Ito, ``{Analytic study of the Migdal-Kadanoff recursion formula},''
  \href{http://dx.doi.org/10.1007/BF01468144}{{\em Commun.Math.Phys.}
  {\bfseries 95} (1984) 247}.

\bibitem{schulz}
V.~Muller and J.~Schiemann, ``{ Convergence of Migdal-Kadanoff Iteraions in
  Nonabelian Lattice Gauge Models },''
  \href{http://dx.doi.org/10.1007/BF01221221}{{\em Commun.Math.Phys.}
  {\bfseries 97} (1985) 605}.

\bibitem{levin}
M.~Levin and C.~Nave, ``{Tensor Renormalization Group Approach to
  Two-Dimensional Classical Lattice Models},''
  \href{http://dx.doi.org/10.1103/PhysRevLett.99.120601}{{\em Physical Review
  Letters} {\bfseries 99} no.~12, (2007) 120601},
  \href{http://arxiv.org/abs/arXiv:cond-mat/0611687}{{\ttfamily
  arXiv:cond-mat/0611687}}.

\bibitem{bd1205}
B.~Dittrich, ``{From the discrete to the continuous: Towards a cylindrically
  consistent dynamics},''
\href{http://arxiv.org/abs/1205.6127}{{\ttfamily arXiv:1205.6127 [gr-qc]}}.
%%CITATION = ARXIV:1205.6127;%%.

\bibitem{sebastian}
B.~Dittrich and S.~Steinhaus, ``{Path integral measure and triangulation
  independence in discrete gravity},''
  \href{http://dx.doi.org/10.1103/PhysRevD.85.044032}{{\em Phys.Rev.}
  {\bfseries D85} (2012) 044032},
\href{http://arxiv.org/abs/1110.6866}{{\ttfamily arXiv:1110.6866 [gr-qc]}}.
%%CITATION = ARXIV:1110.6866;%%.

\bibitem{he}
B.~Bahr, B.~Dittrich, and S.~He, ``{Coarse graining free theories with gauge
  symmetries: the linearized case},''
  \href{http://dx.doi.org/10.1088/1367-2630/13/4/045009}{{\em New J.Phys.}
  {\bfseries 13} (2011) 045009},
  \href{http://arxiv.org/abs/1011.3667}{{\ttfamily arXiv:1011.3667 [gr-qc]}}.

\end{thebibliography}\endgroup

}

\end{document}